\documentclass[twocolumn 
]{aastex631}
\pdfoutput=1

\usepackage{CJK}
\usepackage{xspace}
\usepackage{booktabs}
\usepackage{lmodern}
\usepackage{slantsc}
\usepackage{comment}

\newcommand{\teff}{$T_{\mathrm{eff}}$}
\newcommand{\logg}{$\log(g)$}
\newcommand{\msun}{M$_\odot$}
\newcommand{\sample}{258}

\received{August 16, 2024}
\revised{September 20, 2024}
\accepted{October 7, 2024}

\submitjournal{ApJ}

\shorttitle{Missing Metals on Massive White Dwarfs}
\shortauthors{Ould Rouis et al.}

\graphicspath{{./}{figures/}}

\begin{document}

\begin{CJK*}{UTF8}{gbsn}

\title{Constraints on Remnant Planetary Systems as a Function of Main-Sequence Mass with HST/COS}


\author[0009-0002-6065-3292]{Lou~Baya~Ould~Rouis}
\email{lbor@bu.edu}
\affiliation{Department of Astronomy \& Institute for Astrophysical Research Boston University Boston, MA 02215, USA}

\author[0000-0001-5941-2286]{J.~J.~Hermes}
\affiliation{Department of Astronomy \& Institute for Astrophysical Research Boston University Boston, MA 02215, USA}

\author[0000-0002-2761-3005]{Boris~T.~G\"ansicke}
\affiliation{Department of Physics, University of Warwick, Coventry, CV4 7AL, UK}

\author[0000-0002-0801-8745]{Snehalata~Sahu}
\affiliation{Department of Physics, University of Warwick, Coventry, CV4 7AL, UK}

\author[0000-0002-6164-6978]{Detlev~Koester}
\affiliation{Institut f\"ur Theoretische Physik und Astrophysik, Universit\"at Kiel, 24098 Kiel, Germany}

\author[0000-0001-9873-0121]{P.-E.~Tremblay}
\affiliation{Department of Physics, University of Warwick, Coventry, CV4 7AL, UK}

\author[0000-0001-8014-6162]{Dimitri~Veras}
\affiliation{Department of Physics, University of Warwick, Coventry, CV4 7AL, UK}
\affiliation{Centre for Exoplanets and Habitability, University of Warwick, Coventry CV4 7AL, UK}
\affiliation{Centre for Space Domain Awareness, University of Warwick, Coventry CV4 7AL, UK}

\author[0000-0003-1748-602X]{Jay~Farihi}
\affiliation{Department of Physics \& Astronomy, University College London, Gower Street, London WC1E 6BT, UK}

\author[0000-0003-3868-1123]{Tyler~M.~Heintz}
\affiliation{Department of Astronomy \& Institute for Astrophysical Research Boston University Boston, MA 02215, USA}

\author[0000-0002-6428-4378]{Nicola~Pietro~Gentile Fusillo}
\affiliation{Department of Physics, Universit\`a degli Studi di Trieste, I-34127 Trieste, Italy}

\author[0000-0003-3786-3486]{Seth~Redfield}
\affiliation{Astronomy Department and Van Vleck Observatory, Wesleyan University, Middletown, CT 06459, USA}


\begin{abstract}

As the descendants of stars with masses less than 8\,M$_{\odot}$ on the main sequence, white dwarfs provide a unique way to constrain planetary occurrence around intermediate-mass stars (spectral types BAF) that are otherwise difficult to measure with radial-velocity or transit surveys. We update the analysis of more than 250 ultraviolet spectra of hot ($13{,}000$\,K\,$<$\,\teff\,$<$\,$30{,}000$\,K), young (less than $800$\,Myr) white dwarfs collected by the Hubble Space Telescope, which reveals that more than 40\% of all white dwarfs show photospheric silicon and sometimes carbon, signpost for the presence of remnant planetary systems. 
However, the fraction of white dwarfs with metals significantly decreases for massive white dwarfs (M$_{\rm WD}~>$~0.8\,M$_{\odot}$), descendants of stars with masses greater than 3.5\,M$_{\odot}$ on the main sequence, as just $11^{+6}_{-4}$\% exhibit metal pollution. In contrast, $44\pm6$\% of a subset of white dwarfs (M$\rm _{WD}~<$~0.7\,M$_{\odot}$) unbiased by the effects of radiative levitation are actively accreting planetary debris.
While the population of massive white dwarfs is expected to be influenced by the outcome of binary evolution, we do not find merger remnants to broadly affect our sample. We connect our measured occurrence rates of metal pollution on massive white dwarfs to empirical constraints into planetary formation and survival around stars with masses greater than 3.5\,M$_{\odot}$ on the main sequence.

\end{abstract} 



\section{Introduction} \label{sec:intro}

Recent surveys suggest that our Galaxy is full of exoplanets.
Every star of spectral type K through M in our Galaxy is predicted to host at least one planet within 10~au \citep{2012Natur.481..167C}, and most AFGK stars are expected to host a planet on a orbit under 85-d \citep{2013ApJ...766...81F}. Despite the progress in exoplanet demographics, only two of the more than 5500 confirmed exoplanets orbit stars with a mass greater than 3.5\,\msun\ on the main sequence (after curating the most reliable host masses on the NASA Exoplanet Archive, as of 2024~August, \citealt{2013PASP..125..989A}). Both planetary systems, $\rm \mu^2$ Sco b and b Centauri (AB)b, were discovered by direct imaging \citep{2021Natur.600..231J,2022A&A...664A...9S}.

The lack of constraints in the presence of exoplanets around the more massive stars of our Galaxy can be attributed to sensitivity biases of the most common exoplanet detection methods. In particular, radial velocity is less sensitive for A-type stars and more massive main-sequence stars, which have fewer-to-no narrow spectral lines, larger rotational velocities, as well as more surface variations \citep{2002AN....323..392H}. Transit surveys are most efficient for large planets around small stars, as transit depths are directly proportional to the host-star radius. 

This shortfall can be addressed by moving further along the stellar evolution of late B and A-type main-sequence stars. Planets with sufficiently wide orbits (over a few au) are predicted to survive the giant-branch phase of their host star \citep{2012ApJ...761..121M}, which results in opportunities to search for exoplanets around giants and white dwarfs. First-ascent giants, which have sharp spectral lines, have been studied using precise radial velocities, and long-period planets have been detected around retired A stars (e.g., \citealt{2007ApJ...665..785J}). White dwarfs, the final phase of low- to intermediate-mass stellar evolution (with initial masses under 8\,\msun), are an effective tool to fill in the gaps and investigate planetary system survival. There have been detections of planets around white dwarfs in transit surveys \citep{2020Natur.585..363V}, microlensing surveys  \citep{2021Natur.598..272B}, as well as candidate surviving wide giants planets with direct imaging in JWST \citep{2024ApJ...962L..32M}.

A significant fraction of white dwarfs also offer indirect evidence of remnant planetary systems that can constrain planetary system evolution and survival. The presence of remnant planetary systems around white dwarfs was first inferred from photospheric traces of heavy elements \citep{2003ApJ...596..477Z, 2005A&A...432.1025K}. The strong surface gravity of white dwarfs causes elements heavier than hydrogen or helium to sink below the photosphere on short timescales, orders of magnitude shorter than their cooling ages (e.g., \citealt{1979ApJ...231..826F}, \citealt{2006A&A...453.1051K}). Heavy elements in white dwarfs atmospheres result from the accretion of scattered, tidally disrupted planetesimals onto the star, also resulting in warm debris disks often detected in the infrared (see review by \citealt{2016NewAR..71....9F}). 

The consensus interpretation holds that metal pollution implies the presence of at least one planet and a reservoir of planetesimals \citep{2002ApJ...572..556D, 2016RSOS....350571V}. Observationally, both debris disks and their transits illuminate this model of accretion \citep{2015Natur.526..546V, 2016ApJ...816L..22X, 2017ApJ...839...42R, 2021ApJ...917...41V, 2021ApJ...912..125G, 2022MNRAS.511.1647F}. Some dynamically scattered planetary materials reach the Roche radius of the white dwarf, where tidally disrupted fragments collide until sublimation is possible and metal gas is accreted onto the white dwarf photosphere \citep{2018MNRAS.480.2784V}.

Measuring the abundance ratios of this metal pollution gives insight into the composition of remnant rocky material around the star, and hence offers exceptional insight into the chemical composition of exoplanets and asteroid belt analogs \citep{2007ApJ...671..872Z, 2021Eleme..17..241X}. White dwarf abundances have shed light on 23 elements detected from metal pollution, and have been found to be dominated by rock-forming elements (alike bulk Earth), typically O, Mg, Si, and Fe \citep{2014ApJ...783...79X, 2019MNRAS.490..202S}.

Metal occurrence rates for white dwarfs in the ultraviolet (UV) with spectra obtained by the Cosmic Origin Spectrograph (COS) on the Hubble Space Telescope (HST) first estimated that $27-56$\% of white dwarfs show traces of heavy elements and thus have remnant planetary systems \citep{2014A&A...566A..34K}. Here we extend those constraints on metal pollution fractions among white dwarfs using a far UV spectroscopic survey of hot, young, white dwarfs with HST/COS. We build on the work presented in \citet{2014A&A...566A..34K}, but now analyze a three times larger sample. In Section~\ref{sec:sample}, we introduce the HST/COS sample analyzed here. In Section~\ref{sec:parameters}, we describe the physical parameters for the sample, including white dwarf mass determinations and the connection to progenitor zero-age-main-sequence (ZAMS) mass, and we describe the methods of spectroscopic metal line detections. In Section~\ref{sec:rates}, we present the measured metal pollution fractions and the implications on remnant planetary system occurrence around white dwarfs. In Section~\ref{sec:discussion}, we conclude with a discussion of our results and any observational biases that may affect the lack of massive metal polluted white dwarfs. 


\section{Sample Selection \& Observations} \label{sec:sample}

Our study focuses on white dwarfs observed in the far UV range with HST/COS through Snapshot Programs (PI: G\"ansicke) since 2010, covering data taken from Cycles 18 to 29. Snapshot proposals allow for short observations of objects selected to optimize the HST observing schedule. All HST/COS targets were observed with the G130M grating, which covers 1150\,\AA\ to 1450\,\AA\ at resolving power R\,$\approx 40{,}000$. Exposure times ranged from 400\,s to 2000\,s, with a median of 1200\,s. We obtained reduced spectra from MAST, after use of the CALCOS calibration pipeline version 5.1 \citep{2008SPIE.7014E..6GK, 2022cosd.book..5.1S}. 

Before Gaia massively increased the number of known white dwarf candidates \citep{2019MNRAS.482.4570G, 2021MNRAS.508.3877G}, most targets in this study were selected using the Palomar Green Survey in the northern hemisphere, and the ESO SN Ia Progenitor survey (SPY) survey in the southern hemisphere. Stars were required to have sufficient flux in the UV (over 5~$\times~10^{-14}$ erg cm$^{-2}$ s$^{-1}$ \AA$^{-1}$) in order to achieve a signal-to-noise ratio above 15 for exposures under 2000\,s. Targets were selected with \teff\ under $40{,}000$\,K to assume LTE conditions, while limiting effects of radiative levitation, which can strongly impact the atmospheric composition of white dwarfs with \teff\ above $50{,}000$\,K \citep{1995ApJS...99..189C, 1995ApJ...454..429C}. Additionally, targets were selected with \teff\ above about $15{,}000$\,K, as convection zones become significant around $12{,}000$\,K for hydrogen-rich white dwarfs \citep{1991ApJ...371..719M}, and convection can dredge up metals from deeper layers (e.g., \citealt{2020NatAs...4..663H}).  

All DA white dwarfs in the Snapshot program up to 2023~August were collected and homogeneously analyzed by \citet{2023MNRAS.526.5800S}. 
We exclude DB white dwarfs, whose convection zones become significant at \teff~$\simeq$~$30{,}000$\,K. We also exclude post-main sequence, common-envelope binary systems, magnetic white dwarfs, extremely-low-mass white dwarfs, and mixed atmosphere white dwarfs. These may have complicated evolutionary paths, or in some cases may not arise from single-star evolution. We also exclude WD0525+526, a 1.2\,\msun\ white dwarf with high transverse velocity (see Section~\ref{sec:kinematics}). Based on its location on a color-magnitude diagram, WD0525+526 is a Q-branch merger candidate \citep{2019ApJ...886..100C}. This object shows photospheric detection of carbon that may not come from an extrinsic source (Sahu et al., in prep.); it is likely instead a DAQ \citep{2020NatAs...4..663H, 2024ApJ...965..159K}.

We also exclude binary systems with projected separation under 100\,au (from the wide binary catalog by \citealt{2021MNRAS.506.2269E}), and unresolved binary candidates that may have interacted in the past (see more details in Table~4 of \citealt{2023MNRAS.526.5800S}). 
A total of 35 resolved wide binaries were kept in the sample; 27/35 have wide main-sequence companions. 

Out of the 311 white dwarfs presented in \citet{2023MNRAS.526.5800S}, our selection cuts and applied white dwarf mass limits described in Section~\ref{sec:mass} reduce our sample to \sample\ white dwarfs, presented in Table~\ref{table}.

\section{White Dwarf Parameters}\label{sec:parameters}

\subsection{Atmospheric Parameters}

\begin{figure*}
  \centering
  \includegraphics[width=\textwidth]{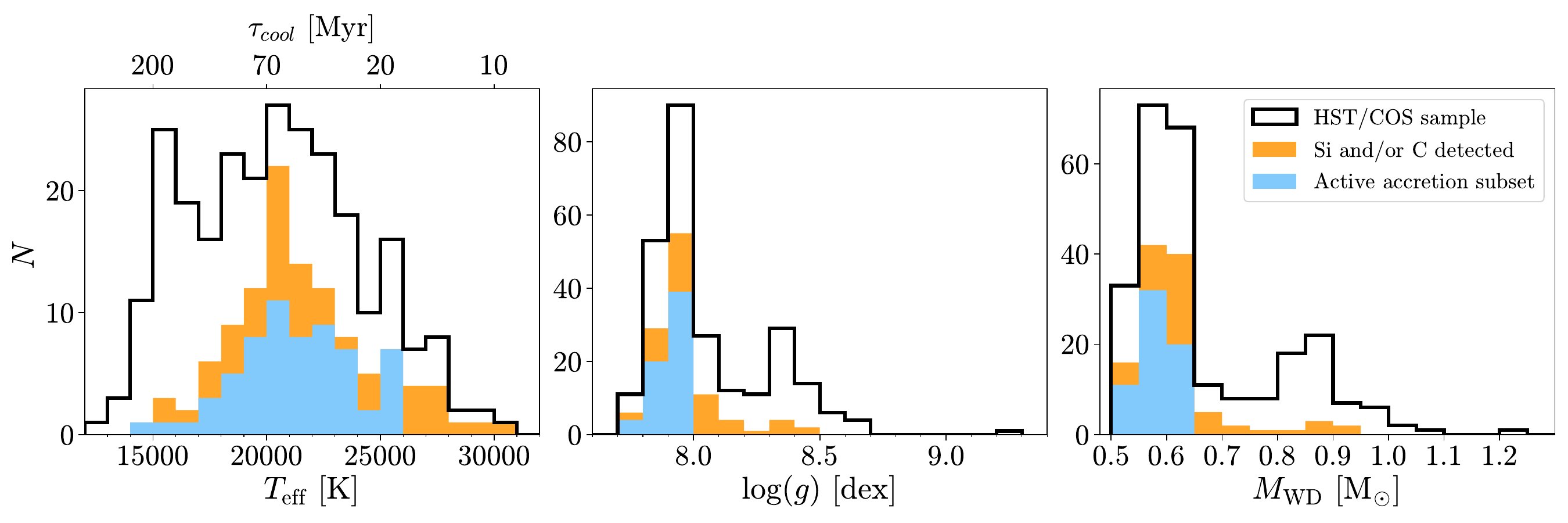}
  \caption{Atmospheric parameters and white dwarf masses of our sample of \sample\ objects observed in the FUV HST/COS Snapshot survey. We highlight the white dwarfs exhibiting some metal pollution, where either photospheric Si and sometimes C absorption lines are observed (orange histogram). The blue histogram shows a subset of white dwarfs polluted by C in a temperature range where radiative levitation effects are inefficient, selected to have an upper mass limit of roughly 0.7\,\msun (see Section~\ref{sec:radlev}). Representative white dwarf cooling ages for a fixed \logg = 8.0 are shown in the top x-axis of the left panel.}\label{fig:params}
\end{figure*}

We compared optical spectroscopic fits from the literature \citep{2009A&A...498..517K, 2011MNRAS.417.1210G, 2011ApJ...730..128T} with UV spectroscopic fits for the HST/COS sample \citep{2023MNRAS.526.5800S}, as well as Gaia photometric fits \citep{2021MNRAS.508.3877G}. There is an unresolved systematic offset between optical and UV atmospheric parameters (see \citealt{2023MNRAS.526.5800S} for discussion). The measured white dwarf masses from UV spectra appear lower by 0.05\,\msun\ and 0.03\,\msun\ than measured from optical spectra and photometry, respectively. Mass and radius determinations from optical photometry are well tested (e.g., \citealt{2019ApJ...876...67B}). 
Therefore we use the hydrogen-dominated Gaia photometric fits from \citet{2021MNRAS.508.3877G} to obtain values for effective temperature, \teff, and surface gravity, \logg. We find a median \teff\ of $20{,}340$\,K and a median $\log(g\,[\mathrm{cm}\,\mathrm{s}^{-2}])$ of 7.97. The full distribution of adopted atmospheric parameters is shown in the black histogram of the first two panels of Figure~\ref{fig:params}. Using the broad bandpass filters of Gaia for these photometric fits, we do not expect significant differences on the mass estimate when using a DA model to fit a DAZ white dwarf (e.g., \citealt{2011MNRAS.417.1210G}).

\subsection{White Dwarf and Progenitor Masses}\label{sec:mass}

We use the {\tt WD\_models} python package from Sihao Cheng\footnote{\url{https://github.com/SihaoCheng/WD_models}} to calculate white dwarf masses and cooling ages of our sample based on \teff\ and \logg, by interpolating existing atmospheric models. We use the CO core, thick-H atmosphere models from \citet{2020ApJ...901...93B}. Since Cycle 25 of the Snapshot program focused on preferentially selecting bright, massive white dwarfs, we do not expect the mass distribution of our sample to follow the prediction from the 100\,pc sample presented in \citet{2021MNRAS.502.4972K}. The mass distribution of our sample shows an over-representation of massive white dwarfs compared to the expected mass distribution of field white dwarfs. 

We assume single-star evolution for our objects, which implies that only white dwarfs above a characteristic mass and below a characteristic total age are considered valid within the age of the Universe. As such, we remove from the sample 40 DAs with masses $\rm M_{WD}~<~0.5$\,\msun. Our sample of \sample\ DA has a median mass of 0.608\,\msun, with a median uncertainty of 0.018\,\msun\ on the mass determination. The full mass distribution is shown in the black histogram of the third panel of Figure~\ref{fig:params}.

\begin{figure}[b]
  \centering
  \includegraphics[width=0.45\textwidth]{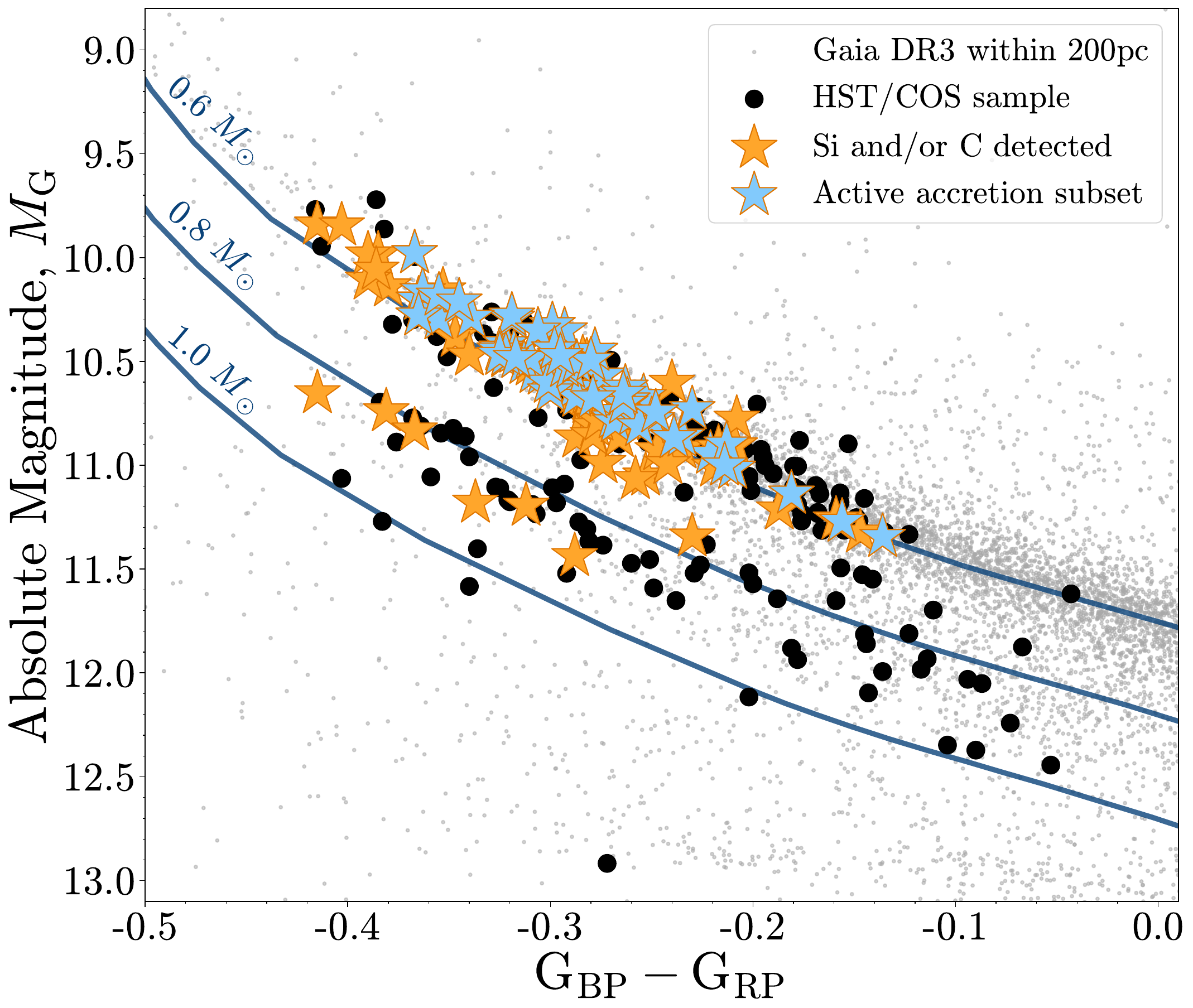}
  \caption{Color-magnitude diagram of the white dwarfs in our HST/COS sample (black points). The orange stars are the white dwarfs exhibiting metal pollution from Si and sometimes C, and we highlight the active accretion subset inefficient to the effects of radiative levitation as stars filled in light blue (see Section~\ref{sec:radlev}). We include all white dwarfs within 200\,pc in Gaia DR3 \citep{2021MNRAS.508.3877G} as gray points as a reference. The solid blue lines show cooling tracks (0.6\,\msun, 0.8\,\msun, 1.0\,\msun) from \citet{2020ApJ...901...93B}, where white dwarfs of a given mass cool from the top left to the bottom right.}\label{fig:CMD}
\end{figure}

Figure~\ref{fig:CMD} shows our HST/COS sample on a color-magnitude diagram compared to the white dwarfs in Gaia within 200\,pc, with representative cooling tracks for 0.6\,\msun, 0.8\,\msun, and 1.0\,\msun\ masses \citep{2020ApJ...901...93B}. White dwarfs cool monotonically along a track from the top left to the bottom right of Figure~\ref{fig:CMD}.

To connect the white dwarfs of our sample to their main sequence progenitors, we use the initial-final mass relation (IFMR) of \cite{2022ApJ...934..148H}. Constraining the lower limits of the IFMR remains a challenge, as there are few direct measurements of low mass white dwarfs in nearby clusters. We set an upper limit progenitor mass of $\rm M_{ZAMS}~<~1$\,\msun\ for white dwarfs with $\rm M_{WD}~<~0.58$\,\msun\ (lower limit for reliable masses per \citealt{2022ApJ...934..148H}). For reference, when comparing a few IFMR from the literature, we find the progenitors of massive white dwarf at 0.8\,\msun\ to be of mass 3.5\,\msun\ from \citet{2022ApJ...934..148H}, 3.4\,\msun\ from \citet{2018ApJ...866...21C}, 3.6\,\msun\ from \citet{2018ApJ...860L..17E}, and 3.7\,\msun\ from \citet{2024MNRAS.527.3602C}. We then assess total derived ages for the objects of our sample using MIST isochrones\footnote{\url{https://waps.cfa.harvard.edu/MIST/index.html}} (evolutionary tracks with $v$/$v_{\rm{crit}}$~=~0.0, [Fe/H]~=~0.0) \citep{2016ApJS..222....8D, 2016ApJ...823..102C} to get the total time spent on the main sequence. The total age is the sum of the time spent on the main sequence and the white dwarf cooling age.

\subsection{Photospheric Metal Detection}

\begin{figure}[t]
  \centering
  {\includegraphics[width=0.45\textwidth]{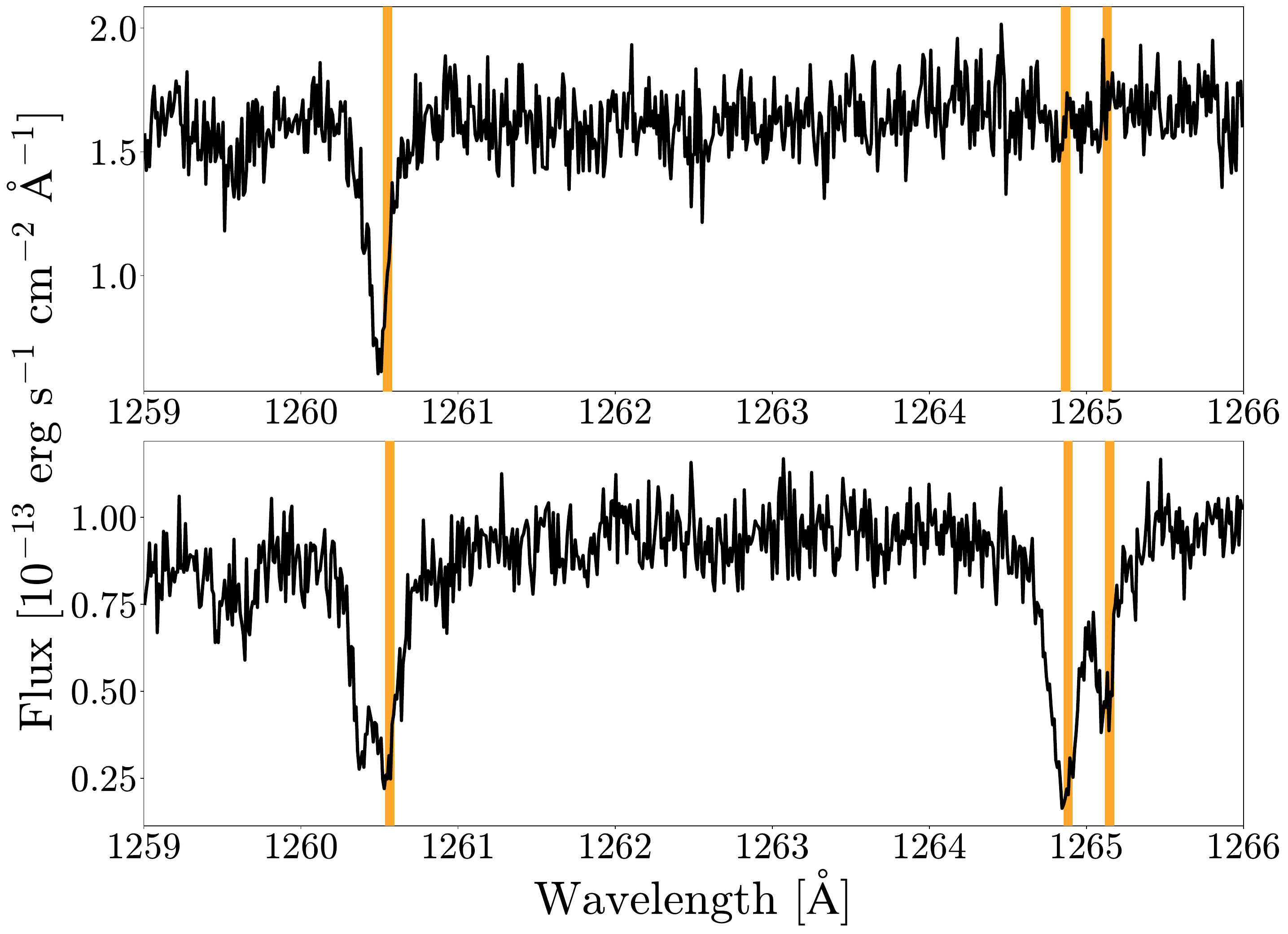}}
  \caption{Representative far UV spectra, illustrated by two objects of our sample, in the wavelengths range where three transitions of Si\,{\sc ii} can be found (marked as orange lines at wavelengths of shifted from rest based on an estimate of the gravitational redshift given the white dwarf mass). The top panel, APASSJ083857.48$-$214611.0 ($\rm M_{WD}$ = 0.58\,\msun), shows no Si pollution and likely only ISM absorption. The bottom panel, WD1953$-$715 ($\rm M_{WD}$ = 0.63\,\msun, RV = +24.9 km s$^{-1}$), features Si pollution in the photosphere, as well as an intervening ISM line near 1260.4\,\AA. } \label{fig:spectra}
\end{figure}

In order to infer metal pollution from white dwarf spectra, we seek absorption lines from heavy elements beyond the broad hydrogen Lyman-$\alpha$ line at 1215.7\,\AA\ expected to dominate for DA white dwarfs. Our UV spectra are typically sensitive to metal pollution detections above log(Si/H) = $-$7.5 \citep{2010AIPC.1273..394C, 2014A&A...566A..34K}, or 10$^5$\,g\,s$^{-1}$ assuming bulk Earth abundances \citep{2003TrGeo...2..547M}. In the UV, Si is expected to best represent the total abundance of metals in the photosphere, as it both has the strongest resonance lines in the UV and also represents a large fraction of inner solar system materials \citep{2012MNRAS.424..333G}. For \teff\ in the range $12{,}000-30{,}000$\,K, we can expect ionized silicon of species Si\,{\sc ii} and Si\,{\sc iii}. The rest wavelength of associated absorption lines are 1260.422\,\AA, 1264.738\,\AA, and 1265.002\,\AA\ for Si\,{\sc ii} and 1294.545\,\AA, 1296.726\,\AA, 1298.892\,\AA, 1298.946\,\AA, and 1301.149\,\AA\ for Si\,{\sc iii}. 
Figure~\ref{fig:spectra} shows two representative HST/COS spectra centered on the Si\,{\sc ii} region. The bottom panel exhibits some metal pollution, while the top panel only exhibits interstellar medium (ISM) absorption. 

We also investigate the presence of carbon absorption lines, with expected species C\,{\sc ii} and C\,{\sc iii} with rest wavelengths 1334.530\,\AA, 1335.660\,\AA, and 1335.708\,\AA\ for C\,{\sc ii} and 1174.930\,\AA, 1175.260\,\AA, 1175.590\,\AA, 1175.710\,\AA, 1175.987\,\AA, and 1176.370\,\AA\ for C\,{\sc iii}. 
In our sample, resonance interstellar medium lines are expected, such as at 1260.422\,\AA\ for Si\,{\sc ii} and 1334.530\,\AA\ for C\,{\sc ii}. Based on equivalent widths, \citet{2014A&A...566A..34K} determine a column densities ranging from 10$^{12}$ to 10$^{13}$ cm$^{-2}$, which are typical for the local insterstellar cloud \citep{2004ApJ...613.1004R}. 

We are able to differentiate ISM from photospheric absorption in a few ways. Photospheric lines are characterized by a sizeable (over 30\,km\,s$^{-1}$) gravitational redshift from rest wavelength, which we can roughly predict based on the white dwarf mass. We flag white dwarf spectra with absorption lines which could be explained by ISM lines, and remove any doubtful example from the metal polluted sample. ISM absorption is almost solely evident in ground-state transitions, so we can remove any doubt of interstellar origin for Si lines when we observe the line at 1265\,\AA, which is always photospheric \citep{2012ApJ...750...69J}, or when C\,{\sc iii} lines are observed, as these excited state are not present in the ISM. 

We inspect our \sample\ HST/COS spectra, presented in Table~\ref{table}, and identify absorption lines coherent with externally sourced pollution from Si and C (see columns $\rm{MP_{Si}}$, $\rm{MP_{C}}$). 

\section{White Dwarf Metal Occurrence} \label{sec:rates}

\subsection{Rates of Si and C Detection}

\begin{figure*}[t]
  \centering
  \includegraphics[width=0.85\textwidth]{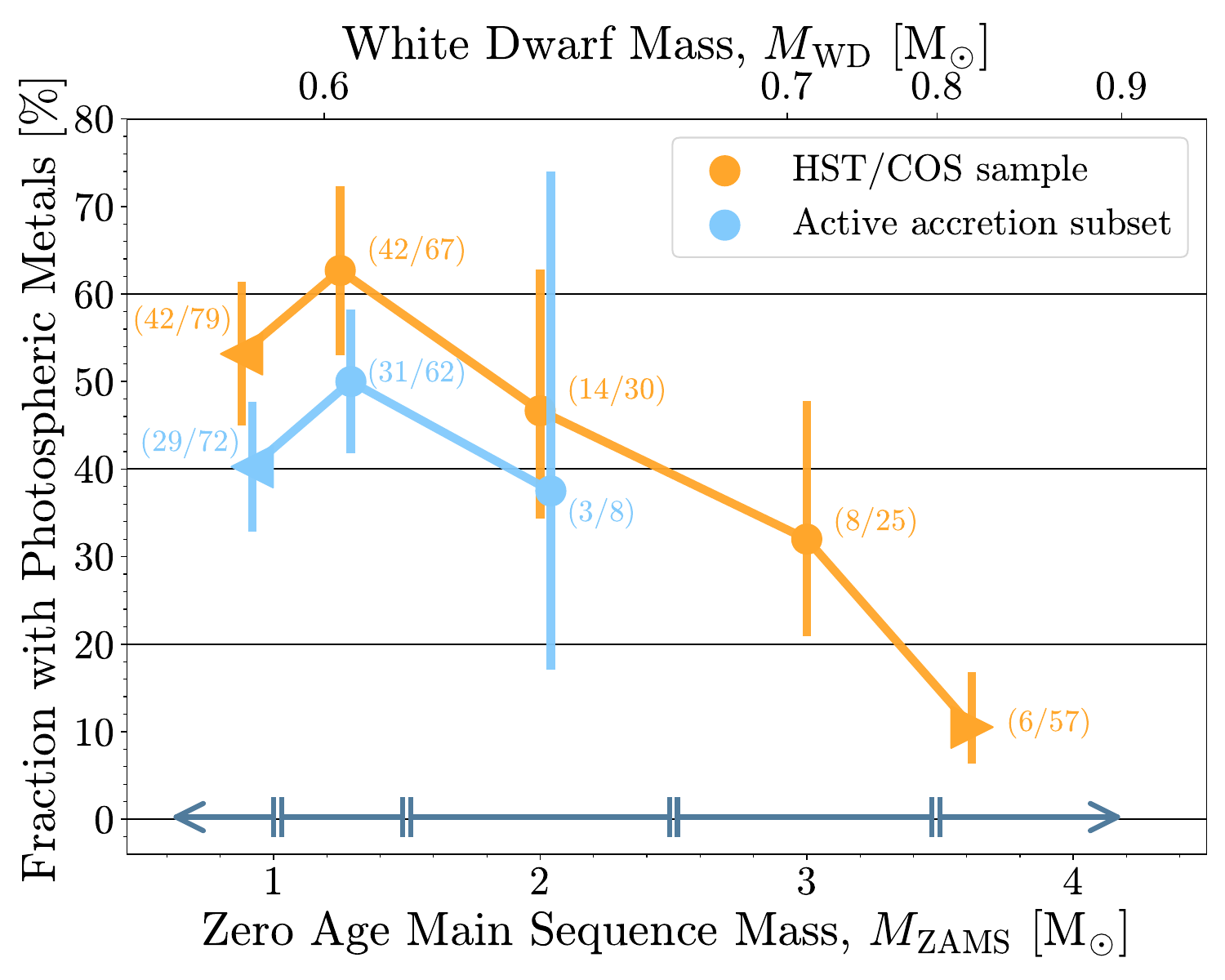}
  \caption{Fraction of white dwarfs showing photospheric metals in our sample (orange) as a function of white dwarf mass (top x-axis label) and zero-age-main-sequence progenitor mass (bottom x-axis label). The dark blue ranges at the bottom delimit the mass bin sizes. Overall, metals are detected in $43\pm4$\% of the full sample across all masses (orange). Just $11^{+6}_{-4}$\% of massive white dwarfs ($\rm M_{WD}$~$>$~0.8\,\msun\ with progenitors $\rm{M_{ZAMS}}$~$>$~3.5\,\msun) show metal pollution. In contrast, $53\pm5$\% of canonical-mass white dwarfs ($\rm M_{WD}$~$<$~0.8\,\msun) show Si and sometimes C lines in the UV. We also plot the fraction of photospheric metals for white dwarfs insensitive to radiative levitation from a subset described in Section~\ref{sec:radlev} in light blue; $44\pm6$\% of this subset is actively accreting planetary debris. White dwarfs with metal lines at velocities which could be explained by ISM contamination are not included in the metal polluted counts.}\label{fig:mp}
\end{figure*}

Within our sample of \sample\ white dwarfs, we find that 112 ($43\pm4$\%) exhibit metal pollution from Si and sometimes C. While this number reflects the fraction of metal polluted white dwarfs in our full HST/COS sample, our sample is heterogenous, and its mass distribution is unrepresentative of field white dwarfs (biased to include more higher-mass white dwarfs, as described in Section~\ref{sec:parameters}). Understanding the $43\pm4$\% frequency relies on the underlying mass distribution of the sample. When excluding the massive white dwarfs, we find that canonical mass white dwarfs ($\rm M_{WD}~<~0.8$\,\msun, from progenitors $\rm M_{ZAMS}~<~3.5$\,\msun) show $53\pm5$\% metal pollution from Si and sometimes C. However, for massive white dwarfs ($\rm M_{WD}~>~0.8$\,\msun), we find just $11^{+6}_{-4}$\% exhibit metal pollution. 

Figure~\ref{fig:mp} illustrates our main result: the photospheric metals fraction as a function of white dwarf mass and corresponding progenitor ZAMS mass. We plot the distribution across different mass bins, chosen to be close to equally spread across the mass range of the sample, to reflect the overall distribution of DAZ as a function of mass. The mass bins cover progenitor mass ranges under 1\,\msun, 1.0 to 1.5\,\msun, 1.5 to 2.5\,\msun, 2.5 to 3.5\,\msun, and over 3.5\,\msun. All numerical fractions are made explicit in the figure, and uncertainties are estimated with Poisson statistics for small numbers \citep{1986ApJ...303..336G}.

The most massive metal-polluted white dwarf in our sample is  WD1459+347 (0.92\,\msun), which shows both Si and possibly C. For $\rm M_{ZAMS}~>5$\,\msun\ progenitors (roughly $\rm M_{WD}$~$>$~0.95\,\msun), no metal pollution is detected (0/10). Thus, we set 1$\sigma$ upper limits of 18\% on white dwarfs with $\rm M_{WD}~>~$0.95\,\msun\ to harbor remnant planetary systems. This is similar in scale to the lack of IR excess in massive white dwarfs determined by \citet{cheng2024}.

Although we have increased the sample size by a factor of three compared to \citet{2014A&A...566A..34K}, we observe similar occurrence rates of metal accretion. Out of the 85 DA white dwarfs they studied, 48 showed photospheric metals, but only 23 were confirmed to be actively accreting; only 1/12 massive white dwarfs showed metals \citep{2014A&A...566A..34K}. They reported equilibrium abundances for all identified DAZ based on diffusion fluxes for the elements, assuming steady state conditions between accretion and diffusion. A complete analysis of the detected element abundance ratios in this larger sample will be presented in future work. 

As white dwarfs in the temperature range covered by the HST/COS sample do not have appreciable convection zones affecting diffusion timescales, we initially assumed that any measured abundances would reflect the active accretion from an external source onto the white dwarf. For the median \teff\ and \logg\ of our HST/COS sample, we predict diffusion timescales of less than a week for both Si and C, while the median cooling age of the sample is tens of millions of years. As diffusion timescales vary with \logg, we expect metals in the photosphere of massive white dwarfs to sink somewhat faster below detection limits given a fixed accretion rate \citep{1986ApJS...61..177P, 2017ASPC..509....3D}. Accurately quantifying the impact of this effect relies on abundance calculations, which will be addressed in future work.

However, radiative levitation is an important effect that can support elements for longer than predicted from diffusion alone, and preferentially acts on lower mass white dwarfs. Radiative levitation effects could alter derived accretion rates and confuse our interpretation of recent or current accretion. We describe how we overcome the potential bias against detection for more massive white dwarfs in the next section.

\subsection{Overcoming Radiative Levitation Timescale Bias}\label{sec:radlev}

Traces of heavy elements can be levitated, counteracting gravitational settling through bound-bound absorption at the surface of white dwarfs above roughly $20{,}000$\,K, powered from the acceleration of individual ions by radiation pressure \citep{1979A&A....80...79V, 1995ApJ...454..429C}. The amount by which an ion can be supported by radiative levitation in the photosphere longer than diffusion depends on the radiative energy flux in the atmosphere, local pressure, and line absorption cross section of the ion. Section~7 of \citet{2014A&A...566A..34K} details the derivation of a grid of models to represent radiative levitation support for each heavy element based on \teff\ (above $10{,}000$\,K) and \logg\ (for values 7.0, 7.5, 8.0, and 8.5), with constant diffusion flux at all layers, assuming homogeneous accretion distributed on surface of star. 

It is important to note that even if radiative levitation is operating for some white dwarfs in our sample, we still require an external source for the metals to be seen in the white dwarf photosphere. Radiative levitation models compiled by \citet{1995ApJ...454..429C} over wider temperature ranges ($20{,}000$\,K to $100{,}000$\,K) confirm that we can theoretically always assume the presence of Si in our sample to originate from an external source at the white dwarf stage because abundances explained by radiative levitation drop below detection levels (log[Si/H] $<$ $-$8.0 even for the lowest surface gravitiy, \logg~=~7.0) at roughly $70{,}000$\,K. White dwarfs in our sample have cooled for a median of 40\,Myr (and 135\,Myr for the oldest white dwarf in our sample) since the abundance of Si supported in the photosphere by radiative levitation was above detection levels, meaning any metal would not have been supported from beyond that point. The polluted sample has a median white dwarf cooling age of 44\,Myr. Thus, most of the $43\pm4$\% of white dwarfs with Si have accreted planetary material within the last 40\,Myr, and all at some point in their cooling as a white dwarf. The accreted metals are interpreted as the signpost of remnant planetary system, indicating the presence of surviving bodies such as planets, asteroids, comets, or other debris \citep{2016RSOS....350571V, 2023MNRAS.519.6257V}.

\citet{2014A&A...566A..34K}, in their first analysis of the HST/COS metal rates, reported that 27\% of white dwarfs (23/85) must be currently accreting because the measured abundances exceed the expected photospheric abundance of that element that can be supported by radiative levitation for the appropriate \teff\ and \logg. In order to claim that an object is undergoing active accretion, \citet{2014A&A...566A..34K} compared the equivalent widths obtained from the modeled synthetic spectra to the observed equivalent widths, and chose the more conservative limits (the observed equivalent width must be a factor of four or larger that of the predicted equivalent width for current accretion to be trusted). \citet{2014A&A...566A..34K} found 23 of the 48 white dwarfs with Si detected to be actively accreting, and 14 out of the 19 objects with C detected to be actively accreting.
Interestingly, HE0416$-$1034 reveals a Si abundance who could be explained by radiative levitation alone, while its C abundance would imply some active accretion. 

While about half of 85 objects published in 2014 with Si absorption lines could be explained by radiative levitation alone, it was only a quarter for objects with C absorption lines. Further, we find that the abundances of the homogeneous C models presented in Figure~6 of \citet{2014A&A...566A..34K} at some point all drop below our detection level of log[C/H]~$<-8$, creating a temperature range in \teff\ unaffected by radiative levitation, such that any detected abundance would imply active accretion. 

We create a corresponding sub-sample, which we refer to as the active accretion subset. This subset includes only objects with \logg~$<7.5$ and \teff~$<19{,}000$\,K, \logg~$<7.75$ and \teff~$<23{,}000$\,K and \logg~$<8.0$ and \teff~$<26{,}000$\,K. The resulting sub-sample is composed of 144 objects, whose parameters distributions can be seen in Figures~\ref{fig:params},~\ref{fig:CMD} in light blue, and whose metal pollution fraction can be seen in Figure~\ref{fig:mp} in light blue. This creates a lower-mass subset of objects all with $\rm M_{WD}<$~0.7\,\msun\ that are unlikely to have major radiative levitation effects.

We detect C lines in 63/142 objects of this active accretion subset, and determine that $44\pm6$\% of white dwarfs with $\rm M_{WD}~<$~0.7\,\msun\ are actively accreting from an external source. 

Radiative levitation support is negligible for the highest \logg, with the Si homogeneous models abundances equal to the lower limit of detection for \logg~=~8.25 and the C homogeneous models abundances under the detection limits for all temperatures for \logg~$>$~8.25 \citep{2014A&A...566A..34K}. All massive white dwarfs in our \teff\ range have \logg~$>$~8.25, so we can assume any detection of Si on a massive target implies active accretion.

The active accretion subset allows us to compare metal pollution occurrence rates across all masses, unbiased by the effects of radiative levitation. We detect metals in $44\pm6$\% of white dwarfs with $\rm M_{WD}~<~$0.7\,\msun\ and $11^{+6}_{-4}$\% of white dwarfs with $\rm M_{WD}~>~$0.8\,\msun. We measure a significant, 4$\sigma$ difference in active accretion from remnant planetary system between canonical-mass and massive white dwarfs. 

\section{Hypotheses for lack of massive DAZ}\label{sec:discussion}

The white dwarfs of the HST/COS sample were selected to provide unbiased insight into remnant planetary system occurrence. Here, we explore different hypothesis for the statistically significant deficit of massive white dwarfs compared to canonical-mass white dwarfs that persists after accounting for radiative levitation. 
 
\subsection{Testing for Merger-Remnant Contamination}

Inferring planetary occurrence rates on the main sequence from white dwarfs relies on the assumption that no mechanism preferentially eliminates planetary systems throughout stellar evolution. There is strong evidence that close double white dwarf systems may merge within a Hubble time \citep{1984ApJ...277..355W,2018MNRAS.476.2584M}, which can lead to Type Ia supernovae, or create a single white dwarf more massive than the individual counterparts \citep{2009A&A...500.1193L}. Based on binary population synthesis calculations, between $10-30$\% of all single white dwarfs, and between $30-50$\% of massive white dwarfs ($\rm M_{WD}>0.8$\,\msun) are formed through binary mergers \citep{2020A&A...636A..31T}, whose violent history may make the survival of planetary bodies unlikely \citep{2007ApJ...661.1192V}. \citet{2014A&A...566A..34K} hypothesized that the lack of metal polluted massive white dwarfs results primarily from merger contamination. Signposts of mergers include strong magnetism, rapid rotation, or atypical kinematics \citep{2023MNRAS.518.2341K}. We explore how we can constrain each effect within our HST/COS sample.

\subsubsection{No Evidence of Strong Magnetism}

A few percent of single white dwarfs harbor strong magnetic fields (over 1\,MG) that can be identified in low-resolution spectra (e.g., \citealt{2013MNRAS.429.2934K}). While the specific causes for strongly magnetic white dwarfs is still an open question, binarity and double white dwarf mergers are expected to generate strong fields in white dwarfs, especially hotter than $10{,}000$\,K \citep{2021MNRAS.507.5902B}. Magnetic white dwarfs are characterized by Zeeman-split spectral features, which can reveal fields of strengths above 2\,MG in low-resolution spectra (see review by \citealt{2015SSRv..191..111F}). 

Our HST/COS sample does not include any known strongly magnetic hydrogen-atmosphere white dwarfs, as we intentionally removed any DAH in our HST/COS sample. While inspecting our spectra for metal pollution detections, we note no signs of Zeeman splitting for any source to a limit of fields roughly over 500\,kG. Based on this, we do not see any strongly magnetic white dwarfs that could be merger remnant candidates in our sample. 

\subsubsection{No Evidence of Rapid Rotation}

The additional angular momentum imparted from stellar merger may cause the resulting white dwarf to rapidly rotate \citep{2021ApJ...906...53S}. While single white dwarfs are expected to rotate with periods around 1\,d \citep{2017ApJS..232...23H}, rapidly rotating white dwarfs are characterized by periods under 20\,min \citep{2021Natur.595...39C, 2023MNRAS.518.2341K}.

We explore photometric variability for our sample by analyzing periodograms from the Transiting Exoplanet Survey Satellite (TESS; \citealt{2015JATIS...1a4003R}). We inspect TESS 20-second- and 2-min-cadence PDCSAP light curves to flag any variability in 228 of our hot DAs (88\% of our sample). 
For the 30 targets that do not have TESS data, 26 have data from the Zwicky Transient Facility (ZTF; \citealt{2019PASP..131a8002B}). We investigate the photometric variability of the remaining four with the All-Sky Automated Survey for SuperNovae (ASAS-SN; \citealt{2014ApJ...788...48S}). The significance thresholds of TESS light curves for our sample (taken as 5$\times$ the mean amplitude of the full periodogram) have a median amplitude of 0.25\%, which is below the observed amplitude reported for the vast majority of spotted white dwarfs with rapid rotations (e.g., \citealt{2020ApJ...894...19R, 2021Natur.595...39C, 2021ApJ...923L...6K, 2022AJ....164..131W, 2024arXiv240704827J}). This implies that non-detection of variability in our sample can give sufficient insight on overall merger remnant contamination. Some targets could show lower-amplitude signals, so we cannot rule out all rapid rotators.
Given the large pixel scale and likely contamination of our sources, we run the TESS-Localize tool \citep{2023AJ....165..141H} to confirm the origin of any detected periodicities in TESS. No periodicities are detected in ZTF or ASAS-SN.

After inspecting all light curves and associated maximum significant periods, we find 18 objects presenting some variability in TESS with a maximum significant period under 1\,d. However, after scrutinizing the origin of the signal with TESS-localize, we find just five targets with an amplitude exceeding 5$\times$ the mean amplitude of the full periodogram with a over 5\% likelihood the variability is coming from the white dwarf. We do not claim with certainty these five are variable (they require higher-resolution follow-up), but include their analysis for completeness.

The five variable candidates are WDJ094755.68$-$ 231234.10, PG1641+388, WDJ180230.44+803951.14, WDJ175352.16+330622.62, and HS0507+0434A. Their respective periods and relative amplitudes in the 2-min-cadence light curves are 6.91428 hr (2.25\%), 3.16942 hr (0.13\%), 3.63507 hr (0.19 \%), 35.98280 min (0.12\%), and 5.92926 min (0.33\%). The last object, HS0507+0434A, is in a wide (over 900\,au) orbit with a well-studied pulsating white dwarf (HS0507+0434B; \citealt{1998AA...330..277J, 2002AA...388..219K}) that is contaminating the TESS photometry, since it is less than 18\,arcsec away.

Our four new candidate variables all have masses $\rm M_{WD}~<$~0.7\,\msun, and are not included in our massive white dwarf subset. Our photometric variability analysis does not find evidence for any rapidly rotating massive white dwarfs in our sample, so we find no evidence for merger contamination from this test of our sample.

\begin{figure*}%
    \centering
    {\includegraphics[width=0.9\textwidth]{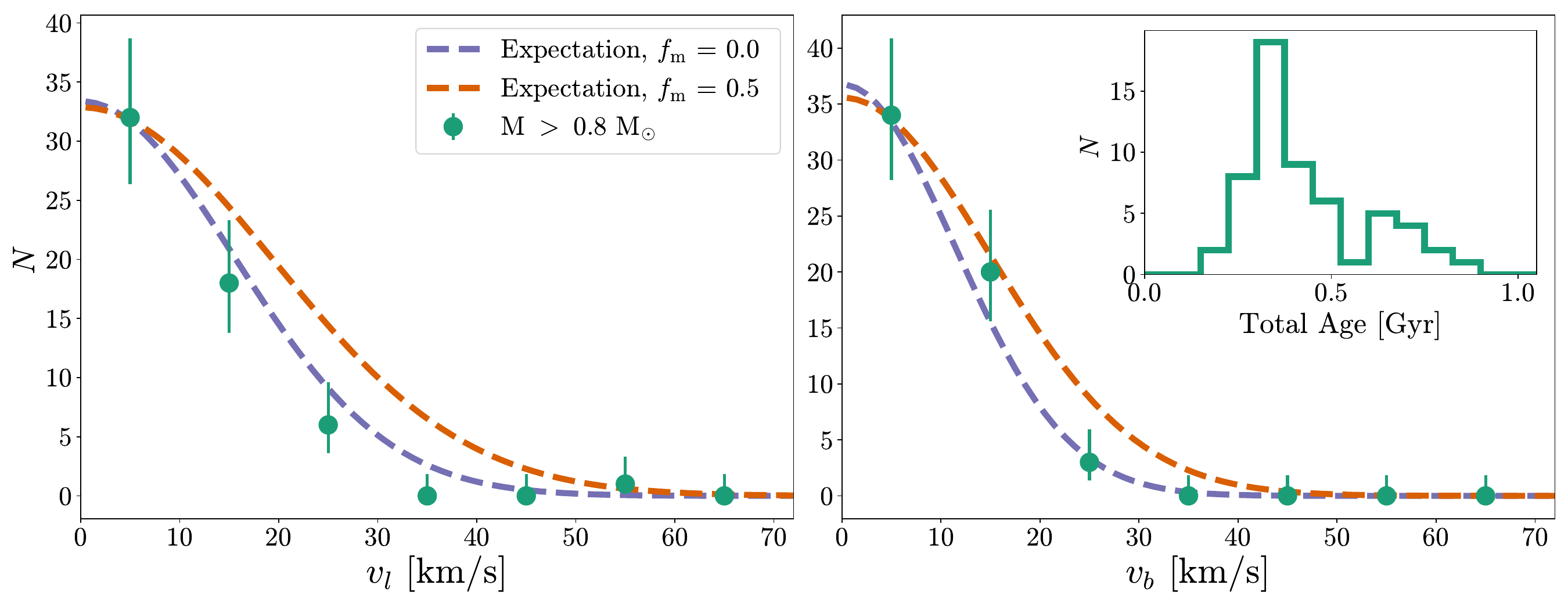}}
    \caption{Transverse velocity distribution of the massive white dwarfs in our sample ($\rm M_{WD}~>$~0.8\,${M_\odot}$, green points) in Galactic coordinates $(l,b)$ compared to expected velocity distributions derived from the total age of the massive white dwarfs in our sample (inset histogram). The green points shows the velocity distributions based on observed proper motions, binned at 10\,km~s$^{-1}$. The derived total white dwarf ages assume some time spent on the main sequence based on an initial-to-final-mass relation. Massive white dwarfs are expected to be younger than 1\,Gyr and to have slow transverse velocities; kinematic outliers are characteristic of merger remnant contamination \citep{2020ApJ...891..160C}. The purple dashed line shows an expected velocity distribution assuming no merger contamination, and the red dashed line assumes 50\% merger contamination (see Section~\ref{sec:kinematics}). We do not see significant kinematic evidence of merger contamination among the 57 massive white dwarfs in our sample.}%
    \label{fig:vel_dist}%
\end{figure*}
    
\subsubsection{No Evidence of Atypical Kinematics}\label{sec:kinematics}

Transverse velocities also allow for identification of merger remnants. Older stars get more stirred up within the Galaxy, causing transverse velocity distributions to follow a theoretical age-velocity dispersion relation \citep{1985ApJ...294..674C}. Because of the age reset produced by a merger, white dwarf merger products will have a higher velocity dispersion than expected \citep{2019ApJ...886..100C}. The massive white dwarfs of our sample should be young, with total ages less than 1\,Gyr; hence, we expect their kinematics to be relatively slow. Crystallisation has not yet kicked in, so we do not expect distillation delays for the large majority of this sample \citep{2019ApJ...886..100C, 2024Natur.627..286B}. Merger delays depend on the type of binary pair involved in the event, as it impacts how significant the age reset generated is \citep{2020A&A...636A..31T}. Individual merger products should show anomalous kinematics; for example, WD0525+526, the hot, 1.2\,\msun\ merger candidate we removed from the sample as discussed in Section~\ref{sec:sample}, has an extremely high longitudinal transverse velocity: $v_l = 116.4$\,km~s$^{-1}$, due to a combination of merger delay and distillation delay.

We are able to measure transverse velocities in Galactic coordinates thanks to precise proper motion measurements from Gaia \citep{2023A&A...674A...1G}. We report values for ($v_l$, $v_b$) in Table~\ref{table} based on methods described in \citet{2020ApJ...891..160C}, corrected for the solar motion around the Galaxy, as well as for asymmetric drift \citep{1998gaas.book.....B}.  Figure~\ref{fig:vel_dist} shows the transverse velocity distribution in Galactic coordinates for the massive white dwarfs ($\rm M_{WD}~>$~0.8\,\msun) of our sample, alongside the total derived age of massive white dwarfs.
Outliers from the massive white dwarf sample are likely merger remnants.

We compute the expected distribution for a null merger fraction, $f_{\rm m} = 0.0$, derived from the age-velocity dispersion relation based on the median age of the massive sample (inset histogram in Figure~\ref{fig:vel_dist}). Following \citet{2019ApJ...886..100C}, a second dashed line shows the expectation distribution for a merger fraction of 50\%, $f_{\rm m} = 0.5$, computed by adding a delay-time distribution of white dwarfs with ages delayed randomly between 0 and 5 Gyr to compare to the population in $f_{\rm m} = 0.0$. The single-star kinematic expectation has a reduced $\chi^2$ fit to the observations that is more than a factor of 2.3 better than the $f_{\rm m} = 0.5$ expectation.

With just 57 massive white dwarfs, it is hard to draw a more precise statistical test given the relatively small numbers of our sample. Still, our kinematic distribution analysis does not find evidence for a large subset of massive white dwarfs with atypical velocities. We see no evidence for significant age resets caused from merger contamination based on this test of our sample.

\subsection{Differences in planets around massive stars}

Our search for merger remnants within our HST/COS sample does not indicate that violent interactions between stars are likely to explain the lack of massive metal polluted white dwarfs. We explore instead possible insights that can be drawn into the formation and evolution of planets around stars with masses $\rm M_{ZAMS}~>$~3.5\,\msun.

\subsubsection{Differences in planetary survival}

Since the observed binary fraction of massive stars ($\rm M_{ZAMS}~>~3.5$\,\msun) exceeds 30\% \citep{2017ApJS..230...15M}, it is somewhat surprising we do not see much evidence for merger contamination in our sample. 
It is likely that our merger contamination analysis is not as sensitive to mergers taking place prior to the white dwarf stage. Roughly 60\% of mergers occur before either component is a white dwarf, which results in short merger delays \citep{2020A&A...636A..31T}. While we intentionally exclude close binary systems from our analysis, if the pair merged long enough before becoming a white dwarf, the age reset may be negligible compared to the white dwarf total age. Some planetary systems may be lost in mergers that do not leave an observational signature.

The observed companion fraction inside 100\,au for mid-B stars is roughly 25\%, compared to roughly 15\% for a FGK stars \citep{2023ASPC..534..275O}. Thus, the difference in binary fraction for the progenitors of the white dwarfs of our study is unlikely to fully explain the factor of four decrease in metals around massive white dwarfs. 

It is worth considering if planets exist around more massive stars on the main sequence but are lost en route to the white dwarf stage. 
A more massive ZAMS progenitors will lose a larger percentage of its initial mass, and faster, than a lower-mass progenitor (e.g., \citealt{2008ApJ...676..594K}). For example, a 3~\msun\ star will lose at least 75\% of its mass in just a few Myr, while a 1~\msun\ star will lose 50\% of its mass over the span of several hundred Myr \citep{2000MNRAS.315..543H}. \citet{2011MNRAS.417.2104V} found that as the mass of the star increases, for a fixed orbital distance, planetary ejection probability from dynamical instabilities increases. This implies that massive white dwarf progenitors may preferentially see orbiting planets ejected during giant branch phases, impacting planetary survival as a function of mass.

Closer-in to the star, critical engulfment distances rely on an equilibrium of mass loss and tidal interactions. Engulfment limits increase by roughly 1 au for every extra solar mass, since more massive stars reach larger radii as giants \citep{2012ApJ...761..121M}. Based on the dynamics of engulfment and ejection, we may assume that massive stars have a narrower orbital parameter space for stable surviving planets.
Yet, massive planets (roughly 10~$\rm M_{Jup}$) are predicted to sometimes survive planetary engulfment at the common envelope stage, either intact or through fragmentation into smaller planets \citep{2012ApJ...759L..30P, 2021MNRAS.501..676L, 2023ApJ...954..176Y}. This gives a new meaning to critical engulfment distances.

Still, the survival of planetary systems around the most massive white dwarf progenitors is possible. \citet{2020MNRAS.493..765V} established limits on required planet separation when the $6-8$\,\msun\ host star leaves the main sequence. Intact planets would need to reside at $a~>~$3~au and $a~>~$6~au for terrestrial and gas giant planets, respectively.
Non-intact minor planets and pebbles, which become disrupted debris, can also reach the white dwarf stage (see \citealt{2020MNRAS.493..765V} for required separation distances as a function of pebbles radii). Thermal radiation forces cause asteroids to undergo semimajor axis drift (the Yarkovsky effect) and spin-vector modifications (YORP spin-up), which lead to altered orbits and tearing apart, respectively  \citep{2006AREPS..34..157B}.

There may be some differences in planetary survival as a function of mass, but both major and minor planets are still expected to reach the white dwarf stage, even for $6-8$\,\msun\ stars. Based on the evolutionary pathways required for white dwarf pollution, we find no definitive reasoning for differences in planetary survival to explain the lack of metals accreting onto massive white dwarfs.

\subsubsection{Differences in planetary architecture}

Though what we know about planets around massive host stars is widely unconstrained by the small number of observations, it is worth considering that the orbital distance or preferential mass of surviving planets could disfavor white dwarf pollution. For example, the most massive main-sequence stars with planets host them on very wide orbits: $\rm \mu^2$ Sco b, and b Centauri (AB)b, were discovered around a 9\,\msun\ star with a projected orbit of 290~au \citep{2022A&A...664A...9S}, and a binary pair of stars with total mass 6--10\,\msun\, with a projected orbit of 550~au  \citep{2021Natur.600..231J}, respectively.

There is no evidence for preferential white dwarf age for accretion to begin. We find signatures of remnant planetary systems around young white dwarfs that have been cooling for $\rm \tau_{cool}~<$~100 Myr (e.g., \citealt{2012MNRAS.424..333G}). We expect accretion to continue on steadily throughout white dwarf cooling, sustained up to at least 1\,Gyr \citep{2022MNRAS.510.1059B}, though there are hints it abates slowly over time \citep{2018MNRAS.477...93H}.

The timescale for scattering events is expected to depend on the orbital distance to the perturber. For Keplerian orbits, a planet at 500~au should interact with asteroids and debris at least two order of magnitudes less frequently then a planet at 15~au. 
Surviving planets which reside at more distant orbits may scatter debris on much less frequent timescale. If it is not an observational selection bias and if massive stars preferentially harbor planets at more distant orbits, it could create a bias against seeing metals in the photospheres of massive white dwarfs as frequently as in lower-mass stars.

\subsubsection{Differences in planetary formation}

The final hypothesis calls back to the early stages of stellar evolution, as planets may simply not form as frequently around massive stars. \citet{2020MNRAS.499.1890M} studied white dwarfs in the Gaia 40\,pc sample and also identify a significant scarcity of higher mass ($\rm M_{WD}\,\gtrsim$ 0.7\,\msun) DAZ compared to DA white dwarfs. They speculate that planet formation may be less efficient at the higher mass stars. Infrared excesses from dust disks also appear rare around massive white dwarfs, suggesting giant planet occurrence rates are very low around main-sequence stars $\rm M_{ZAMS}~>~$\,3\,\msun\ \citep{cheng2024}.

All planetary formation processes rely on circumstellar, protoplanetary disks, and the mass of a star should be proportional to the mass of its associated disk. More massive stars are expected to host more massive disks, which create more massive planets \citep{2010PASP..122..905J}. 

Observationally, there are still no constraints on the most likely mass of the planet responsible for the scattering of debris towards the white dwarfs. It is predicted the lowest-mass perturber to drive the accretion onto a white dwarf is about one Luna mass \citep{2023MNRAS.519.6257V}.
On the one hand, if planetary systems contain only low-mass objects, the chances of potential pollutants to escape the host star are lower (e.g. \citealt{2023MNRAS.519.6257V}). For more massive stars whose planetary systems are expected to contain more massive planets, pollutants may escape more frequently. This may leave the highest-mass stars with smaller reservoirs of planetary debris at the white dwarf stage \citep{2021MNRAS.506.1148V}. 

Even with massive disks, the lack of confirmed exoplanets around massive stars has been suspected to be caused by formation-halted and migration-halted scenarios \citep{2021Natur.600..231J}. The rapid dispersion timescales of disks around massive stars from high-energy radiation may cause disk lifetimes to be shorter than giant planets formation timescales \citep{1997Sci...276.1836B}. Planets may not have had enough time to form around the most massive stars, in which case the decrease in massive white dwarfs showing metal pollution is a direct way to constrain the inefficiencies of planet formation around massive main-sequence stars.


\section{Conclusion}\label{sec:summary}

Our study of UV spectra from HST/COS confirms the results presented in \citet{2014A&A...566A..34K} that metal polluted white dwarfs are common. 
When differentiating the sample by mass, we find that $53\pm5$\% of white dwarfs with masses $0.5-0.8$\,\msun, which evolve from AFG stars on the main sequence, show active or recent accretion of metals and thus host remnant planetary systems. While some of these white dwarfs may have metals supported by radiative levitation now, that radiative support was inefficient at some point in its past, meaning that these metal-polluted objects require accretion from planetary debris at some point in their history as a white dwarf.

We find a lack of massive metal polluted white dwarfs, as just $11^{+6}_{-4}$ of white dwarfs with mass $\rm M_{WD}$~$>$~0.8\,\msun\ (evolved from main-sequence B stars) show detectable Si or C in their UV spectra. 
Since radiative levitation may bias us from detecting metals in more massive white dwarfs, we use a active accretion subset insensitive to radiative levitation effects to provide limits on active planetary system accretion for the lower-mass white dwarfs, mitigating biases that may affect the timescales for which we would see pollution in massive white dwarfs. This subset, all with $\rm M_{WD}$~$<$~0.7\,\msun, shows $44\pm6$\% metal pollution. Therefore, we find that massive white dwarfs have a 4.0$\sigma$ lower incidence in the active accretion of remnant planetary systems compared to canonical-mass white dwarfs. 

Based on three observational signposts of merger remnants (magnetism, rapid rotation, and anomalous kinematics), we do not find our massive white dwarf sample to be significantly contaminated by merger remnants. Since we have selected against close binary systems, we do not expect the interactions between stars to have played a dominant factor in the survival of planetary systems in our sample. 

We explore a number of possible reasons why we see at least four times fewer massive white dwarfs with remnant planetary systems, but do not reach a conclusion on the dominant mechanism; it is possible that multiple causes are simultaneously responsible. For example, it is possible that planets are more commonly lost during the late-stage evolution of more massive stars. It is also possible that the disk lifetimes are too short to form planets around massive stars. 

There are additional avenues to explore to see if planetary system formation and survival is different around intermediate-mass main-sequence stars. As this HST/COS program is running low on bright UV white dwarfs targets, we must explore different methods to improve the statistics presented in this work. Massive multiplexed spectroscopic surveys like SDSS-V \citep{2023ApJS..267...44A} and DESI \citep{2023arXiv230606308D} will yield orders of magnitude more spectroscopically identified white dwarfs with photospheric metals. Narrow-band Ca imaging should also significantly increase known metal-rich white dwarfs (e.g., \citealt{2024arXiv240616055L}). Exploring these larger samples as a function of mass should aid in understanding the missing remnant planetary systems around massive white dwarfs, and thus better constrain the true planetary occurrence rate around B-type main sequence stars. 

\section{Acknowledgments}
\begin{acknowledgments}

The authors thank the anonymous referee for the productive and beneficial feedback. We recognize and appreciate Siyi~Xu (许\CJKfamily{bsmi}偲\CJKfamily{gbsn}艺), Sihao~Cheng (程思浩), Joseph~Guidry, Tim~Cunningham, and Kevin~Schlaufman for helpful discussions in the preparation of this manuscript. 
Support for this work was in part provided by NASA TESS Cycle 4 grant 80NSSC22K0737 and Cycle 6 grant 80NSSC24K0878, as well as NASA under Grant No. 80NSSC23K1068.
This research was improved by discussions at the KITP Program ``White Dwarfs as Probes of the Evolution of Planets, Stars, the Milky Way and the Expanding Universe'' supported by National Science Foundation under grant No. NSF PHY-1748958.
This project has received funding from the European Research Council (ERC) under the European Union's Horizon 2020 research and innovation programme, grant agreements 101002408 (MOS100PC) and 101020057 (WDPLANETS).

This research is based on observations made with the NASA/ESA Hubble Space Telescope obtained from the Space Telescope Science Institute, which is operated by the Association of Universities for Research in Astronomy, Inc., under NASA contract NAS 5-26555. These observations are associated with programs \#12169, \#12474, \#13652, \#14077, \#15073, \#16011, and \#16642. Funding was in-part provided by programs \#15073, \#16642, and \#17420.
This work has made use of data from the European Space Agency (ESA) mission Gaia (https://www.cosmos.esa.int/gaia), processed by the Gaia Data Processing and Analysis Consortium (DPAC, https://www.cosmos.esa.int/web/gaia/dpac/consortium). Funding for the DPAC has been provided by national institutions, in particular the institutions participating in the Gaia Multilateral Agreement. \\

The data presented in this article were obtained from the Mikulski Archive for Space Telescopes (MAST) at the Space Telescope
Science Institute. The specific observations analyzed can be accessed at \dataset[DOI]{https://doi.org/10.17909/reeb-8718}.

\end{acknowledgments}


\appendix

We provide in Table~\ref{table} our sample of \sample\, white dwarfs observed with HST/COS, ranked by white dwarf mass in descending order. We report the object name, atmospheric parameters (\teff\, and \logg), white dwarf mass ($\rm M_{WD}$), progenitor main-sequence mass ($\rm M_{ZAMS}$), and white dwarf cooling age ($\tau_{\rm WD}$). The columns MP$_{\rm Si}$ and MP$_{\rm C}$ refer to the detection of photospheric metals in the white dwarf spectra. Finally, we report transverse velocities in Galactic coordinates corrected for the Sun's motion around our Galaxy ($v_l$ and $v_b$).


\begin{longtable}{lccccccccc}
\caption{} \\
\toprule
\textbf{Name} & \teff & \logg &  $\rm M_{WD}$ & $\rm M_{ZAMS}$ &  $\tau_{\rm WD}$ & $\rm{MP_{Si}}$ & $\rm{MP_{C}}$ & $v_l$ & $v_b$ \\

& (K) & (cgs) & (\msun) & (\msun) & (Myr) &  &  & (km\,s$^{-1}$) & (km\,s$^{-1}$) \\
\midrule
\endfirsthead
\caption{(Continued)}\\
\toprule
\textbf{Name} & \teff & \logg &  $\rm M_{WD}$ & $\rm M_{ZAMS}$ &  $\tau_{\rm WD}$ & $\rm{MP_{Si}}$ & $\rm{MP_{C}}$ & $v_l$ & $v_b$ \\
& (K) & (cgs) & (\msun) & (\msun) & (Myr) &  &  & (km\,s$^{-1}$) & (km\,s$^{-1}$) \\
\midrule
\endhead
\midrule
\endfoot
\bottomrule
\endlastfoot
WD1518+636 & 21400  (660) & 9.21  (0.03) & 1.23  (0.01) & 8.0 & 712  (71) & 0 & 0 & 51.77 & 12.8 \\
WD1058$-$129 & 24470  (530) & 8.69  (0.02) & 1.05  (0.01) & 6.2 & 162  (19) & 0 & 0 & 11.88 & 7.22 \\
WD1103+384 & 27540  (1780) & 8.65  (0.09) & 1.03  (0.05) & 6.0 & 100  (44) & 0 & 0 & 8.53 & $-$19.47 \\
WD1102$-$281 & 17460  (400) & 8.65  (0.03) & 1.02  (0.02) & 5.7 & 398  (45) & 0 & 0 & 1.96 & $-$3.22 \\
WD0034$-$602 & 14380  (220) & 8.6  (0.01) & 0.99  (0.01) & 5.5 & 623  (38) & 0 & 0 & $-$15.85 & $-$8.92 \\
WD0730+487 & 14100  (240) & 8.59  (0.01) & 0.98  (0.01) & 5.4 & 647  (45) & 0 & 0 & 21.14 & $-$15.68 \\
HE0308$-$2305 & 24000  (530) & 8.57  (0.03) & 0.98  (0.02) & 5.4 & 134  (17) & 0 & 0 & 10.61 & $-$12.91 \\
WDJ053420.32$-$021431.22 & 27900  (1140) & 8.56  (0.05) & 0.98  (0.03) & 5.3 & 76  (24) & 0 & 0 & 0.53 & 7.4 \\
WD0231$-$054 & 13140  (200) & 8.58  (0.01) & 0.98  (0.01) & 5.3 & 767  (49) & 0 & 0 & 9.55 & 22.41 \\
WD0808+435 & 21740  (640) & 8.53  (0.03) & 0.95  (0.02) & 4.9 & 169  (27) & 0 & 0 & 9.42 & $-$8.38 \\
WD1410+081 & 15610  (280) & 8.52  (0.02) & 0.94  (0.01) & 4.8 & 432  (36) & 0 & 0 & 6.66 & $-$8.37 \\
WD1449+513 & 16730  (450) & 8.49  (0.03) & 0.93  (0.02) & 4.6 & 341  (45) & 0 & 0 & $-$1.98 & $-$1.03 \\
WD1459+347 & 21800  (560) & 8.48  (0.03) & 0.92  (0.02) & 4.6 & 152  (22) & 1 & 1 & 9.92 & 20.38 \\
WDJ080420.96+225017.36 & 19200  (630) & 8.47  (0.04) & 0.91  (0.02) & 4.4 & 221  (36) & 0 & 0 & 2.03 & $-$1.0 \\
WD0701$-$587 & 13440  (230) & 8.48  (0.01) & 0.91  (0.01) & 4.4 & 611  (42) & 0 & 0 & $-$17.9 & 7.27 \\
WD1038+633 & 24250  (510) & 8.45  (0.03) & 0.91  (0.02) & 4.3 & 99  (14) & 1 & 0 & 18.44 & 4.91 \\
WD2039$-$682 & 16420  (270) & 8.46  (0.02) & 0.9  (0.01) & 4.3 & 338  (25) & 0 & 0 & $-$16.7 & $-$0.9 \\
WD2043$-$635 & 25250  (600) & 8.44  (0.03) & 0.9  (0.02) & 4.2 & 80  (15) & 0 & 0 & $-$7.83 & $-$3.73 \\
WDJ174902.45$-$343255.27 & 19070  (500) & 8.44  (0.03) & 0.9  (0.02) & 4.1 & 213  (28) & 0 & 0 & $-$0.92 & 2.41 \\
WDJ191720.56+445239.38 & 23200  (600) & 8.43  (0.03) & 0.89  (0.02) & 4.0 & 109  (18) & 0 & 0 & 7.94 & $-$19.03 \\
WD0743+442 & 15200  (260) & 8.44  (0.02) & 0.89  (0.01) & 4.0 & 405  (31) & 0 & 0 & 22.64 & $-$11.19 \\
WDJ210952.38+650721.93 & 21410  (480) & 8.42  (0.02) & 0.88  (0.02) & 4.0 & 143  (18) & 0 & 0 & 5.65 & 0.17 \\
WD1102+748 & 20030  (440) & 8.42  (0.02) & 0.88  (0.02) & 3.9 & 176  (20) & 0 & 0 & 19.46 & $-$2.6 \\
WD1446+286 & 23270  (490) & 8.41  (0.02) & 0.88  (0.02) & 3.9 & 102  (14) & 0 & 0 & $-$3.75 & $-$6.31 \\
HS2056+0721 & 26660  (700) & 8.4  (0.04) & 0.88  (0.03) & 3.9 & 56  (15) & 0 & 0 & 13.78 & $-$7.81 \\
WD2350$-$248 & 29310  (800) & 8.39  (0.04) & 0.88  (0.03) & 3.9 & 32  (11) & 1 & 1 & $-$7.85 & 17.31 \\
SDSSJ081305.55+140317.4 & 20560  (630) & 8.4  (0.04) & 0.87  (0.03) & 3.8 & 155  (28) & 0 & 0 & 7.17 & $-$9.13 \\
WD2220+133 & 22850  (690) & 8.39  (0.04) & 0.87  (0.02) & 3.8 & 106  (22) & 0 & 0 & 6.53 & $-$6.39 \\
WD1550+183 & 14810  (220) & 8.4  (0.02) & 0.86  (0.01) & 3.8 & 409  (28) & 0 & 0 & 6.55 & 29.79 \\
WDJ030236.65$-$230151.23 & 22880  (570) & 8.39  (0.03) & 0.86  (0.02) & 3.8 & 104  (18) & 0 & 0 & 7.01 & $-$1.52 \\
WD1052+273 & 22500  (480) & 8.39  (0.02) & 0.86  (0.02) & 3.8 & 111  (15) & 1 & 0 & 4.13 & $-$13.78 \\
WD1547+057 & 22770  (620) & 8.38  (0.04) & 0.86  (0.02) & 3.8 & 105  (20) & 0 & 0 & $-$10.78 & 15.34 \\
WDJ004331.10+470134.30 & 21360  (620) & 8.38  (0.03) & 0.86  (0.02) & 3.8 & 132  (23) & 0 & 0 & $-$16.27 & $-$3.12 \\
HS2220+2146B & 19120  (540) & 8.39  (0.04) & 0.86  (0.02) & 3.8 & 190  (29) & 0 & 0 & $-$7.2 & $-$19.15 \\
PG0004+061 & 23500  (630) & 8.38  (0.04) & 0.86  (0.02) & 3.8 & 92  (19) & 0 & 0 & 10.55 & $-$15.61 \\
WD0348+339 & 14010  (220) & 8.39  (0.02) & 0.85  (0.01) & 3.7 & 464  (32) & 0 & 0 & 26.95 & 11.63 \\
WD1130$-$125 & 14080  (230) & 8.39  (0.02) & 0.85  (0.01) & 3.7 & 456  (33) & 0 & 0 & $-$6.43 & $-$12.25 \\
WD1049$-$158 & 19300  (400) & 8.37  (0.02) & 0.85  (0.02) & 3.7 & 181  (20) & 0 & 0 & $-$17.14 & $-$2.93 \\
WD2205+250 & 26630  (550) & 8.36  (0.03) & 0.85  (0.02) & 3.7 & 49  (10) & 1 & 0 & 5.48 & $-$12.91 \\
WDJ220238.75$-$280942.13 & 21100  (470) & 8.36  (0.03) & 0.85  (0.02) & 3.7 & 131  (18) & 0 & 0 & $-$5.25 & 1.64 \\
HE0418$-$1021 & 22940  (670) & 8.36  (0.04) & 0.85  (0.02) & 3.7 & 96  (20) & 0 & 0 & 5.38 & 6.7 \\
WD0406+169 & 15420  (280) & 8.37  (0.02) & 0.84  (0.02) & 3.7 & 346  (31) & 0 & 0 & 18.06 & 18.18 \\
WD1120+439 & 27150  (750) & 8.34  (0.04) & 0.84  (0.03) & 3.7 & 41  (13) & 1 & 0 & 16.25 & $-$4.6 \\
WD0558+165 & 17230  (470) & 8.36  (0.03) & 0.84  (0.02) & 3.7 & 247  (33) & 0 & 0 & $-$1.04 & $-$11.91 \\
HS0400+1451 & 14670  (210) & 8.36  (0.02) & 0.84  (0.01) & 3.6 & 391  (25) & 0 & 0 & 18.98 & 19.06 \\
WD0947+325 & 21860  (540) & 8.34  (0.03) & 0.83  (0.02) & 3.6 & 110  (18) & 0 & 0 & $-$2.07 & $-$4.52 \\
WD1452+553 & 27440  (730) & 8.32  (0.04) & 0.83  (0.03) & 3.6 & 36  (12) & 0 & 0 & 17.72 & 16.3 \\
WD1233$-$164 & 24000  (520) & 8.32  (0.03) & 0.82  (0.02) & 3.6 & 72  (13) & 0 & 0 & $-$16.72 & $-$6.99 \\
APASSJ152827.83$-$251503.0 & 15270  (280) & 8.34  (0.02) & 0.82  (0.02) & 3.6 & 337  (30) & 0 & 0 & $-$19.21 & 4.21 \\
HS2210+2323 & 22580  (720) & 8.32  (0.04) & 0.82  (0.03) & 3.6 & 92  (22) & 0 & 0 & $-$0.01 & $-$0.74 \\
WDJ191429.35$-$544019.71 & 25550  (550) & 8.31  (0.03) & 0.82  (0.02) & 3.6 & 51  (11) & 0 & 0 & $-$1.88 & 2.15 \\
WD1523+322 & 25320  (670) & 8.3  (0.04) & 0.82  (0.02) & 3.6 & 53  (14) & 0 & 0 & 0.61 & 16.07 \\
WD1334$-$160 & 18140  (440) & 8.32  (0.03) & 0.82  (0.02) & 3.6 & 199  (26) & 0 & 0 & $-$23.08 & $-$0.59 \\
WD0232+525 & 17350  (310) & 8.32  (0.02) & 0.82  (0.01) & 3.6 & 228  (20) & 0 & 0 & 22.43 & $-$5.3 \\
WD0052$-$147 & 25760  (550) & 8.29  (0.03) & 0.81  (0.02) & 3.5 & 46  (11) & 0 & 0 & $-$0.92 & $-$6.7 \\
WD2205$-$139 & 24860  (560) & 8.29  (0.03) & 0.81  (0.02) & 3.5 & 56  (12) & 0 & 0 & $-$1.55 & $-$0.12 \\
WD1525+257 & 22140  (510) & 8.29  (0.03) & 0.8  (0.02) & 3.5 & 92  (16) & 0 & 0 & $-$24.23 & $-$15.47 \\
PG1220+234 & 25050  (660) & 8.28  (0.04) & 0.8  (0.03) & 3.5 & 50  (14) & 0 & 0 & $-$4.7 & $-$18.47 \\
WD0922+183 & 24500  (700) & 8.27  (0.05) & 0.8  (0.03) & 3.5 & 56  (17) & 0 & 0 & 5.6 & 14.71 \\
WD2047+372 & 14710  (200) & 8.3  (0.02) & 0.79  (0.01) & 3.5 & 349  (21) & 0 & 0 & 8.95 & 4.25 \\
WDJ204745.04+323922.58 & 18860  (370) & 8.28  (0.02) & 0.79  (0.02) & 3.5 & 163  (18) & 0 & 0 & $-$14.91 & 1.86 \\
WD1911+536 & 17190  (310) & 8.29  (0.02) & 0.79  (0.01) & 3.5 & 220  (20) & 0 & 0 & 9.39 & $-$4.03 \\
HE2238$-$0433 & 16100  (920) & 8.29  (0.08) & 0.79  (0.05) & 3.5 & 268  (81) & 0 & 0 & 13.48 & $-$5.78 \\
WD0556+172 & 18450  (440) & 8.25  (0.03) & 0.77  (0.02) & 3.4 & 165  (23) & 1 & 0 & 5.19 & 2.84 \\
HS2220+2146A & 13710  (280) & 8.24  (0.02) & 0.76  (0.02) & 3.3 & 391  (36) & 0 & 0 & $-$7.71 & $-$19.58 \\
WDJ045219.36+251933.98 & 21140  (440) & 8.17  (0.03) & 0.73  (0.02) & 3.1 & 83  (13) & 0 & 0 & 18.07 & $-$10.82 \\
APASSJ013001.36+263857.4 & 14390  (230) & 8.19  (0.02) & 0.73  (0.01) & 3.1 & 316  (24) & 0 & 0 & $-$25.51 & $-$9.07 \\
WDJ002313.53+475259.55 & 20120  (440) & 8.16  (0.03) & 0.72  (0.02) & 3.1 & 100  (16) & 1 & 0 & 36.38 & 5.53 \\
WD1034+492 & 20460  (490) & 8.14  (0.03) & 0.71  (0.02) & 3.0 & 88  (16) & 1 & 0 & $-$11.99 & 32.67 \\
WDJ101839.84$-$310802.03 & 15510  (250) & 8.15  (0.02) & 0.71  (0.01) & 3.0 & 238  (20) & 0 & 0 & $-$30.49 & $-$2.32 \\
WD0732$-$427 & 15030  (230) & 8.15  (0.02) & 0.7  (0.01) & 3.0 & 261  (20) & 0 & 0 & $-$81.36 & 64.93 \\
WD1919+145 & 15160  (230) & 8.15  (0.02) & 0.7  (0.01) & 3.0 & 253  (19) & 0 & 0 & $-$13.97 & 6.06 \\
WD0321$-$026 & 28730  (2970) & 8.1  (0.19) & 0.7  (0.11) & 2.9 & 12  (19) & 0 & 0 & 31.93 & 5.59 \\
WDJ193124.43+570419.66 & 22700  (510) & 8.12  (0.03) & 0.7  (0.02) & 2.9 & 49  (11) & 0 & 0 & $-$3.71 & $-$2.8 \\
WD0421+162 & 19410  (400) & 8.12  (0.02) & 0.7  (0.02) & 2.9 & 104  (15) & 1 & 0 & 17.42 & 17.57 \\
WD0431+126 & 21150  (440) & 8.11  (0.03) & 0.69  (0.02) & 2.8 & 69  (12) & 1 & 0 & 13.89 & 17.96 \\
WD1104+602 & 18320  (350) & 8.11  (0.02) & 0.69  (0.01) & 2.8 & 127  (15) & 0 & 0 & 34.02 & 9.93 \\
WDJ180230.44+803951.14 & 25450  (600) & 8.07  (0.03) & 0.68  (0.02) & 2.8 & 22  (5) & 0 & 0 & $-$24.48 & 14.32 \\
WDJ072805.02$-$130256.34 & 23160  (490) & 8.07  (0.03) & 0.67  (0.02) & 2.7 & 38  (8) & 0 & 0 & 55.86 & $-$34.23 \\
WDJ182315.21+170639.42 & 21840  (460) & 8.06  (0.03) & 0.66  (0.02) & 2.6 & 51  (10) & 1 & 1 & $-$40.96 & $-$13.95 \\
WDJ175712.24+283957.46 & 19330  (410) & 8.07  (0.03) & 0.66  (0.02) & 2.6 & 92  (15) & 1 & 0 & $-$32.0 & 0.58 \\
WDJ165112.59$-$204106.36 & 20410  (440) & 8.06  (0.03) & 0.66  (0.02) & 2.5 & 71  (13) & 1 & 1 & $-$13.57 & $-$5.09 \\
WDJ081004.00+032926.91 & 19750  (420) & 8.05  (0.03) & 0.65  (0.02) & 2.4 & 80  (14) & 0 & 0 & 2.8 & $-$13.59 \\
WDJ193955.06+093219.39 & 22120  (1190) & 8.04  (0.07) & 0.65  (0.04) & 2.4 & 44  (22) & 0 & 0 & $-$3.27 & $-$9.56 \\
PG1641+388 & 15900  (280) & 8.06  (0.02) & 0.65  (0.01) & 2.4 & 186  (19) & 0 & 0 & 28.3 & $-$2.15 \\
WD2306+124 & 20210  (480) & 8.04  (0.03) & 0.65  (0.02) & 2.4 & 70  (14) & 1 & 0 & $-$4.21 & 10.11 \\
APASSJ081237.87+173700.3 & 16030  (270) & 8.04  (0.02) & 0.64  (0.01) & 2.2 & 177  (17) & 0 & 0 & 17.85 & 14.74 \\
WDJ085102.69$-$615517.65 & 20320  (430) & 8.03  (0.03) & 0.64  (0.02) & 2.2 & 66  (11) & 1 & 0 & $-$6.44 & $-$2.93 \\
HE0414$-$4039 & 21130  (580) & 8.02  (0.04) & 0.64  (0.02) & 2.2 & 54  (13) & 0 & 0 & 39.09 & 30.56 \\
WD0904+391 & 24950  (840) & 8.0  (0.06) & 0.63  (0.03) & 2.1 & 21  (7) & 0 & 0 & 12.21 & 17.27 \\
WDJ175151.11$-$202308.72 & 16470  (420) & 8.03  (0.03) & 0.63  (0.02) & 2.1 & 157  (24) & 1 & 1 & $-$6.19 & $-$1.92 \\
PG0817+386 & 25810  (760) & 7.99  (0.05) & 0.63  (0.03) & 2.1 & 17  (4) & 0 & 0 & 32.98 & $-$42.5 \\
WDJ091918.15$-$473354.38 & 23820  (520) & 8.0  (0.03) & 0.63  (0.02) & 2.1 & 27  (5) & 1 & 1 & 3.73 & 15.44 \\
WD1953$-$715 & 18880  (390) & 8.02  (0.03) & 0.63  (0.02) & 2.0 & 90  (14) & 1 & 1 & $-$54.44 & 13.5 \\
HS0507+0434A & 20590  (430) & 8.01  (0.03) & 0.63  (0.02) & 2.0 & 59  (10) & 1 & 0 & 23.18 & $-$8.3 \\
PG0915+526 & 16140  (330) & 8.02  (0.03) & 0.63  (0.02) & 2.0 & 166  (21) & 0 & 0 & 7.63 & 13.01 \\
WDJ112401.30$-$505938.44 & 16380  (290) & 8.02  (0.02) & 0.63  (0.01) & 2.0 & 157  (17) & 0 & 0 & 35.49 & 2.15 \\
HS0944+1913 & 17080  (320) & 8.02  (0.02) & 0.63  (0.02) & 2.0 & 134  (16) & 0 & 0 & 3.25 & $-$16.81 \\
WD0013$-$241 & 18700  (410) & 8.01  (0.03) & 0.63  (0.02) & 1.9 & 92  (16) & 0 & 0 & $-$26.0 & 58.95 \\
WD2134+218 & 18260  (380) & 8.01  (0.03) & 0.63  (0.02) & 1.9 & 102  (15) & 1 & 0 & $-$24.63 & 23.96 \\
WDJ192034.41$-$471529.44 & 19120  (400) & 8.01  (0.03) & 0.63  (0.02) & 1.9 & 83  (14) & 0 & 0 & $-$10.09 & $-$0.27 \\
WD1325+279 & 20300  (620) & 8.0  (0.04) & 0.62  (0.03) & 1.8 & 62  (17) & 1 & 1 & $-$4.48 & 18.33 \\
WD2018$-$233 & 15320  (280) & 8.02  (0.02) & 0.62  (0.02) & 1.8 & 196  (21) & 0 & 0 & $-$8.84 & $-$7.78 \\
WDJ191810.96+723724.89 & 22950  (550) & 7.98  (0.03) & 0.62  (0.02) & 1.8 & 31  (7) & 0 & 0 & 9.44 & $-$37.22 \\
WD1548+149 & 21600  (490) & 7.99  (0.03) & 0.62  (0.02) & 1.7 & 44  (9) & 1 & 1 & $-$25.89 & $-$5.92 \\
WD1929+011 & 21540  (460) & 7.99  (0.03) & 0.62  (0.02) & 1.7 & 44  (9) & 1 & 1 & $-$18.49 & 21.4 \\
PG1421+318 & 27740  (680) & 7.96  (0.04) & 0.62  (0.02) & 1.7 & 12  (2) & 1 & 1 & $-$17.56 & 20.8 \\
WD0102+095 & 24570  (520) & 7.98  (0.03) & 0.62  (0.02) & 1.7 & 22  (4) & 1 & 0 & $-$21.53 & $-$7.26 \\
WD1258+593 & 15180  (280) & 8.01  (0.02) & 0.62  (0.01) & 1.7 & 200  (20) & 0 & 0 & $-$17.59 & $-$13.7 \\
WD0220+222 & 15320  (340) & 8.01  (0.03) & 0.62  (0.02) & 1.6 & 192  (25) & 0 & 0 & 62.11 & 7.54 \\
WD0307+149 & 21620  (530) & 7.98  (0.03) & 0.62  (0.02) & 1.6 & 42  (10) & 1 & 0 & 5.76 & $-$40.7 \\
WDJ214125.64$-$484953.75 & 15080  (250) & 8.01  (0.02) & 0.62  (0.01) & 1.6 & 202  (18) & 1 & 0 & $-$22.81 & 17.89 \\
WD1527+090 & 21500  (480) & 7.98  (0.03) & 0.61  (0.02) & 1.5 & 43  (9) & 0 & 0 & 5.28 & $-$12.3 \\
WDJ230202.99$-$263048.08 & 19570  (440) & 7.98  (0.03) & 0.61  (0.02) & 1.5 & 70  (13) & 0 & 0 & $-$1.25 & $-$22.41 \\
HS2225+2158 & 26170  (670) & 7.96  (0.04) & 0.61  (0.02) & 1.4 & 15  (3) & 1 & 1 & $-$53.46 & 1.73 \\
WD2021$-$128 & 20550  (470) & 7.98  (0.03) & 0.61  (0.02) & 1.4 & 55  (11) & 0 & 0 & $-$50.65 & 10.67 \\
PG1601+581 & 15560  (260) & 8.0  (0.02) & 0.61  (0.01) & 1.4 & 180  (17) & 1 & 0 & $-$1.39 & 6.85 \\
WDJ083920.71$-$280132.44 & 25310  (550) & 7.96  (0.03) & 0.61  (0.02) & 1.4 & 18  (3) & 1 & 1 & $-$9.56 & $-$5.91 \\
WD1713+695 & 15730  (270) & 8.0  (0.02) & 0.61  (0.01) & 1.4 & 173  (17) & 0 & 0 & $-$48.85 & 25.44 \\
WD0048+202 & 20130  (500) & 7.98  (0.03) & 0.61  (0.02) & 1.4 & 61  (13) & 1 & 0 & 30.4 & $-$8.87 \\
WDJ215229.65+340743.85 & 18910  (380) & 7.98  (0.03) & 0.61  (0.02) & 1.4 & 82  (13) & 0 & 0 & 28.17 & $-$20.91 \\
WDJ184933.96+402015.54 & 17980  (360) & 7.99  (0.02) & 0.61  (0.02) & 1.4 & 103  (14) & 1 & 1 & $-$3.87 & $-$14.58 \\
WD0410+117 & 20760  (430) & 7.98  (0.03) & 0.61  (0.02) & 1.4 & 51  (9) & 1 & 0 & 27.36 & $-$1.17 \\
WD2058+181 & 17860  (370) & 7.99  (0.03) & 0.61  (0.02) & 1.4 & 106  (15) & 1 & 1 & 1.66 & $-$31.63 \\
PG1143+321 & 15730  (280) & 8.0  (0.02) & 0.61  (0.01) & 1.4 & 172  (18) & 0 & 0 & 50.69 & $-$2.25 \\
APASSJ145521.26+565544.3 & 15460  (270) & 8.0  (0.02) & 0.61  (0.01) & 1.4 & 183  (18) & 1 & 1 & 8.34 & 20.51 \\
WDJ230840.77$-$214459.60 & 16050  (300) & 7.99  (0.02) & 0.61  (0.01) & 1.4 & 160  (18) & 0 & 0 & $-$10.39 & $-$43.64 \\
WD2341+322 & 12590  (180) & 8.0  (0.01) & 0.61  (0.01) & 1.3 & 346  (21) & 0 & 0 & $-$27.26 & 2.02 \\
PG1126+384 & 25060  (550) & 7.95  (0.03) & 0.61  (0.02) & 1.3 & 19  (3) & 1 & 1 & 25.56 & $-$20.6 \\
HS2229+2335 & 18500  (540) & 7.98  (0.04) & 0.61  (0.03) & 1.3 & 89  (21) & 1 & 1 & 27.62 & $-$25.13 \\
WD0352+018 & 22240  (570) & 7.96  (0.04) & 0.61  (0.02) & 1.3 & 35  (8) & 1 & 1 & 45.94 & 31.45 \\
WD1257+048 & 22220  (500) & 7.96  (0.03) & 0.61  (0.02) & 1.3 & 35  (7) & 1 & 1 & $-$25.89 & $-$43.67 \\
WD1308$-$301 & 14730  (270) & 7.99  (0.02) & 0.61  (0.02) & 1.3 & 213  (22) & 1 & 1 & $-$35.05 & $-$19.06 \\
WD1647+375 & 22740  (520) & 7.96  (0.03) & 0.61  (0.02) & 1.3 & 30  (6) & 1 & 1 & $-$9.73 & 29.03 \\
WDJ063541.34$-$052430.64 & 21570  (470) & 7.96  (0.03) & 0.61  (0.02) & 1.3 & 41  (8) & 1 & 1 & $-$26.57 & 20.15 \\
WD0854+404 & 22610  (510) & 7.96  (0.03) & 0.61  (0.02) & 1.3 & 31  (6) & 0 & 0 & 73.03 & 17.36 \\
WD0000+171 & 21270  (680) & 7.96  (0.05) & 0.61  (0.03) & 1.3 & 44  (13) & 1 & 0 & 15.9 & $-$38.55 \\
HS0002+1635 & 25680  (730) & 7.94  (0.05) & 0.6  (0.03) & 1.3 & 16  (3) & 1 & 1 & 40.28 & $-$19.05 \\
WD1633+676 & 23620  (960) & 7.95  (0.06) & 0.6  (0.03) & 1.3 & 24  (8) & 0 & 0 & $-$28.17 & 3.72 \\
PG1508+549 & 17090  (390) & 7.98  (0.03) & 0.6  (0.02) & 1.2 & 123  (19) & 0 & 0 & $-$21.42 & 29.37 \\
WDJ172730.92+090051.98 & 20850  (440) & 7.96  (0.03) & 0.6  (0.02) & 1.2 & 48  (9) & 1 & 1 & $-$27.44 & $-$4.56 \\
HS2244+2103 & 23980  (780) & 7.95  (0.06) & 0.6  (0.03) & 1.2 & 23  (6) & 1 & 1 & $-$28.58 & $-$15.95 \\
WDJ094755.68$-$231234.10 & 22930  (600) & 7.95  (0.04) & 0.6  (0.02) & 1.2 & 29  (7) & 1 & 1 & 26.64 & $-$17.97 \\
WD1015+161 & 18380  (470) & 7.97  (0.04) & 0.6  (0.02) & 1.2 & 90  (18) & 1 & 0 & 20.76 & $-$48.15 \\
WDJ015630.05+295532.28 & 15260  (260) & 7.98  (0.02) & 0.6  (0.01) & 1.2 & 185  (18) & 0 & 0 & 19.42 & $-$16.69 \\
WDJ175352.16+330622.62 & 17090  (310) & 7.97  (0.02) & 0.6  (0.01) & 1.2 & 122  (15) & 0 & 0 & $-$9.43 & 14.31 \\
HE1247$-$1130 & 26600  (600) & 7.93  (0.04) & 0.6  (0.02) & 1.2 & 14  (2) & 1 & 1 & $-$0.01 & $-$28.42 \\
WD2319+691 & 20060  (430) & 7.96  (0.03) & 0.6  (0.02) & 1.2 & 58  (10) & 1 & 1 & $-$42.3 & 19.43 \\
WD1005+642 & 19680  (400) & 7.96  (0.02) & 0.6  (0.01) & 1.2 & 64  (10) & 0 & 0 & 25.5 & 7.44 \\
APASSJ225612.94$-$131939.3 & 19650  (650) & 7.96  (0.05) & 0.6  (0.03) & 1.2 & 64  (18) & 1 & 1 & $-$17.74 & $-$14.75 \\
WD0242$-$174 & 20550  (530) & 7.96  (0.04) & 0.6  (0.02) & 1.2 & 51  (11) & 1 & 0 & 15.76 & 9.96 \\
WD1609+044 & 30360  (880) & 7.91  (0.05) & 0.6  (0.03) & 1.2 & 9  (1) & 0 & 1 & $-$27.34 & 7.84 \\
WDJ205314.31$-$001608.81 & 22740  (520) & 7.94  (0.03) & 0.6  (0.02) & 1.2 & 29  (6) & 1 & 1 & $-$38.36 & $-$0.19 \\
HE0403$-$4129 & 22700  (790) & 7.94  (0.05) & 0.6  (0.03) & 1.2 & 30  (9) & 1 & 1 & 11.56 & $-$6.67 \\
SDSSJ170029.93+422452.99 & 23720  (1030) & 7.94  (0.07) & 0.6  (0.04) & 1.2 & 23  (9) & 0 & 0 & $-$11.18 & 17.06 \\
WD2046$-$220 & 23140  (570) & 7.94  (0.04) & 0.6  (0.02) & 1.2 & 27  (6) & 1 & 1 & $-$5.51 & $-$23.4 \\
WDJ081425.47$-$643211.05 & 18340  (360) & 7.96  (0.03) & 0.6  (0.02) & 1.2 & 89  (13) & 0 & 0 & 11.26 & 8.23 \\
WDJ105925.27$-$724409.93 & 19210  (420) & 7.96  (0.03) & 0.6  (0.02) & 1.2 & 71  (12) & 1 & 1 & 1.91 & 20.57 \\
WD1017+125 & 21170  (680) & 7.95  (0.04) & 0.6  (0.03) & 1.1 & 43  (12) & 1 & 1 & 18.04 & $-$12.98 \\
WD1133+293 & 23160  (540) & 7.94  (0.03) & 0.6  (0.02) & 1.1 & 26  (5) & 1 & 1 & 32.14 & $-$1.95 \\
WD1330+473 & 23040  (510) & 7.94  (0.03) & 0.6  (0.02) & 1.1 & 27  (5) & 1 & 1 & 4.72 & 22.25 \\
WDJ073548.24+022423.49 & 22670  (560) & 7.94  (0.03) & 0.6  (0.02) & 1.1 & 29  (6) & 1 & 1 & 15.45 & 2.24 \\
APASSJ202336.88$-$111551.3 & 16160  (290) & 7.96  (0.02) & 0.6  (0.01) & 1.1 & 148  (16) & 1 & 1 & $-$28.31 & $-$22.98 \\
WD1327$-$083 & 14560  (220) & 7.97  (0.02) & 0.59  (0.01) & 1.1 & 213  (17) & 0 & 0 & $-$102.54 & $-$28.71 \\
WD1310$-$305 & 19720  (410) & 7.95  (0.03) & 0.59  (0.02) & 1.1 & 61  (11) & 1 & 1 & $-$2.23 & $-$29.11 \\
HE0416$-$1034 & 25610  (580) & 7.92  (0.04) & 0.59  (0.02) & 1.1 & 16  (2) & 1 & 1 & 28.43 & 32.63 \\
WDJ123456.22+473733.37 & 15230  (250) & 7.96  (0.02) & 0.59  (0.01) & 1.1 & 181  (17) & 0 & 0 & $-$1.93 & 62.29 \\
WD0140$-$392 & 22010  (450) & 7.93  (0.03) & 0.59  (0.02) & 1.1 & 34  (6) & 1 & 0 & $-$26.05 & 43.55 \\
WDJ203838.15$-$332635.20 & 18400  (360) & 7.95  (0.03) & 0.59  (0.01) & 1.1 & 85  (12) & 1 & 1 & $-$39.67 & $-$7.28 \\
WDJ003310.51+474212.39 & 16430  (310) & 7.95  (0.02) & 0.59  (0.01) & 1.1 & 136  (16) & 0 & 0 & $-$2.16 & 7.74 \\
WD1145+187 & 27020  (590) & 7.9  (0.03) & 0.59  (0.02) & 1.0 & 13  (1) & 1 & 1 & 15.74 & $-$12.63 \\
WDJ030146.30+493659.64 & 16350  (300) & 7.95  (0.02) & 0.59  (0.01) & 1.0 & 138  (16) & 0 & 0 & 84.56 & $-$4.6 \\
WDJ150742.03$-$592754.43 & 20490  (500) & 7.93  (0.03) & 0.59  (0.02) & 1.0 & 49  (10) & 0 & 0 & $-$45.17 & $-$10.13 \\
WD1408+323 & 18440  (380) & 7.94  (0.03) & 0.59  (0.01) & 1.0 & 82  (12) & 1 & 1 & $-$85.63 & 40.86 \\
WDJ082532.35$-$072823.21 & 15370  (250) & 7.95  (0.02) & 0.59  (0.01) & 1.0 & 172  (16) & 0 & 0 & 28.09 & $-$8.44 \\
WD1555$-$089 & 14400  (240) & 7.95  (0.02) & 0.59  (0.01) & 1.0 & 214  (20) & 0 & 0 & $-$80.36 & $-$11.06 \\
WDJ191558.47$-$303535.44 & 17120  (340) & 7.94  (0.03) & 0.59  (0.02) & 1.0 & 113  (15) & 0 & 0 & $-$42.79 & $-$3.65 \\
WDJ061000.36+281428.37 & 18270  (390) & 7.93  (0.03) & 0.58  (0.02) & 1.0 & 85  (13) & 0 & 0 & 69.06 & $-$35.01 \\
WDJ003043.68+733738.23 & 20020  (480) & 7.92  (0.03) & 0.58  (0.02) & 1.0 & 54  (11) & 1 & 1 & 43.24 & 17.06 \\
WDJ174127.11$-$650342.07 & 19460  (400) & 7.93  (0.03) & 0.58  (0.02) & $<$\,1.0 & 63  (10) & 1 & 0 & $-$64.05 & 10.87 \\
APASSJ083857.48$-$214611.0 & 21380  (470) & 7.92  (0.03) & 0.58  (0.02) & $<$\,1.0 & 38  (7) & 0 & 0 & 12.2 & $-$0.19 \\
WD1507+220 & 20250  (490) & 7.92  (0.03) & 0.58  (0.02) & $<$\,1.0 & 51  (10) & 1 & 1 & $-$63.14 & 38.0 \\
WD2322$-$181 & 22040  (670) & 7.91  (0.04) & 0.58  (0.02) & $<$\,1.0 & 32  (8) & 1 & 1 & 15.05 & $-$120.86 \\
HE0358$-$5127 & 23280  (610) & 7.9  (0.04) & 0.58  (0.02) & $<$\,1.0 & 24  (5) & 0 & 0 & 11.35 & 31.53 \\
WDJ152310.59+305344.80 & 25560  (560) & 7.89  (0.03) & 0.58  (0.02) & $<$\,1.0 & 16  (2) & 1 & 1 & $-$57.71 & 9.24 \\
PG1513+442 & 27600  (700) & 7.88  (0.04) & 0.58  (0.02) & $<$\,1.0 & 12  (1) & 1 & 1 & $-$13.92 & 64.26 \\
WDJ150156.33+302258.23 & 24280  (500) & 7.89  (0.03) & 0.58  (0.02) & $<$\,1.0 & 20  (3) & 1 & 1 & $-$22.91 & 42.16 \\
WD1914$-$598 & 18880  (390) & 7.92  (0.03) & 0.58  (0.02) & $<$\,1.0 & 71  (11) & 1 & 1 & $-$51.96 & 21.52 \\
WDJ181140.82+282939.19 & 18630  (390) & 7.92  (0.03) & 0.58  (0.02) & $<$\,1.0 & 75  (12) & 0 & 0 & $-$85.58 & $-$16.44 \\
WDJ050824.06+213419.83 & 15770  (300) & 7.93  (0.03) & 0.57  (0.01) & $<$\,1.0 & 150  (18) & 0 & 0 & 38.72 & $-$15.77 \\
PG1113+413 & 25440  (690) & 7.88  (0.04) & 0.57  (0.02) & $<$\,1.0 & 16  (2) & 1 & 1 & 40.66 & 31.25 \\
WD1449+168 & 22070  (580) & 7.9  (0.04) & 0.57  (0.02) & $<$\,1.0 & 31  (7) & 1 & 0 & $-$24.86 & 18.16 \\
WDJ170707.97+222428.34 & 15420  (260) & 7.93  (0.02) & 0.57  (0.01) & $<$\,1.0 & 163  (16) & 0 & 0 & $-$53.79 & 24.34 \\
WD0956+020 & 16550  (410) & 7.92  (0.04) & 0.57  (0.02) & $<$\,1.0 & 124  (22) & 0 & 0 & 50.45 & 2.28 \\
WD0106$-$358 & 28800  (670) & 7.86  (0.04) & 0.57  (0.02) & $<$\,1.0 & 10  (1) & 1 & 1 & 26.75 & 62.96 \\
WD1020$-$207 & 19500  (430) & 7.9  (0.03) & 0.57  (0.02) & $<$\,1.0 & 59  (11) & 1 & 1 & 1.52 & $-$40.81 \\
WDJ124202.13+750845.51 & 19230  (420) & 7.9  (0.03) & 0.57  (0.02) & $<$\,1.0 & 62  (11) & 1 & 1 & 96.43 & 44.04 \\
WDJ170634.56$-$184047.13 & 20100  (520) & 7.9  (0.04) & 0.57  (0.02) & $<$\,1.0 & 50  (11) & 1 & 1 & $-$55.53 & $-$3.8 \\
WD1451+006 & 25200  (570) & 7.87  (0.04) & 0.57  (0.02) & $<$\,1.0 & 17  (2) & 0 & 0 & $-$120.28 & 46.31 \\
WDJ012942.65+422818.11 & 23150  (510) & 7.88  (0.03) & 0.57  (0.02) & $<$\,1.0 & 24  (4) & 0 & 0 & 63.93 & $-$19.6 \\
WD0954+697 & 21660  (590) & 7.89  (0.04) & 0.57  (0.02) & $<$\,1.0 & 34  (7) & 1 & 0 & 34.53 & 55.9 \\
WDJ051002.85$-$003755.65 & 17410  (340) & 7.91  (0.03) & 0.57  (0.01) & $<$\,1.0 & 98  (14) & 1 & 0 & 39.03 & 11.36 \\
WDJ055046.46+261220.67 & 21210  (560) & 7.89  (0.04) & 0.57  (0.02) & $<$\,1.0 & 37  (8) & 0 & 0 & 79.67 & 20.0 \\
WDJ012813.78$-$530011.30 & 16570  (300) & 7.9  (0.02) & 0.56  (0.01) & $<$\,1.0 & 119  (14) & 0 & 0 & $-$55.24 & 57.5 \\
WDJ184157.88+533818.93 & 22200  (530) & 7.87  (0.03) & 0.56  (0.02) & $<$\,1.0 & 29  (5) & 1 & 1 & $-$58.97 & 13.72 \\
WD1314$-$153 & 15660  (280) & 7.9  (0.03) & 0.56  (0.01) & $<$\,1.0 & 148  (17) & 0 & 0 & $-$105.97 & $-$193.24 \\
HS1243+0132 & 19900  (710) & 7.88  (0.06) & 0.56  (0.03) & $<$\,1.0 & 50  (16) & 1 & 1 & $-$33.75 & $-$43.18 \\
WDJ192726.24+100710.03 & 24580  (520) & 7.86  (0.03) & 0.56  (0.02) & $<$\,1.0 & 18  (2) & 1 & 1 & $-$116.09 & $-$1.94 \\
WD0018$-$339 & 20450  (450) & 7.88  (0.03) & 0.56  (0.02) & $<$\,1.0 & 44  (8) & 0 & 0 & $-$89.02 & $-$19.75 \\
WDJ082130.53$-$251140.78 & 20960  (520) & 7.88  (0.03) & 0.56  (0.02) & $<$\,1.0 & 38  (8) & 1 & 1 & 74.73 & 25.38 \\
WD2248$-$504 & 15680  (290) & 7.9  (0.02) & 0.56  (0.01) & $<$\,1.0 & 145  (17) & 0 & 0 & $-$90.51 & 2.04 \\
WDJ184915.07$-$212603.48 & 21640  (580) & 7.87  (0.04) & 0.56  (0.02) & $<$\,1.0 & 32  (7) & 1 & 1 & $-$83.2 & 14.76 \\
WD2143+353 & 23860  (610) & 7.86  (0.04) & 0.56  (0.02) & $<$\,1.0 & 20  (3) & 0 & 0 & 99.46 & $-$33.15 \\
PG1202+309 & 29250  (1570) & 7.83  (0.09) & 0.56  (0.05) & $<$\,1.0 & 9  (2) & 0 & 0 & 90.62 & 11.35 \\
WDJ104017.14$-$655324.81 & 20820  (520) & 7.87  (0.03) & 0.56  (0.02) & $<$\,1.0 & 39  (8) & 1 & 1 & $-$18.24 & $-$30.2 \\
WD1755+194 & 23890  (630) & 7.85  (0.04) & 0.56  (0.02) & $<$\,1.0 & 20  (3) & 1 & 1 & $-$99.13 & $-$1.24 \\
HE0418$-$5326 & 26570  (860) & 7.84  (0.06) & 0.56  (0.03) & $<$\,1.0 & 13  (2) & 1 & 1 & $-$0.34 & 14.38 \\
PG1339+346 & 16620  (600) & 7.89  (0.05) & 0.55  (0.03) & $<$\,1.0 & 114  (28) & 0 & 0 & $-$60.9 & 37.64 \\
HE1518$-$0020 & 15380  (340) & 7.89  (0.03) & 0.55  (0.02) & $<$\,1.0 & 154  (21) & 0 & 0 & $-$100.13 & $-$2.54 \\
WDJ022339.21+510454.25 & 18120  (400) & 7.88  (0.03) & 0.55  (0.02) & $<$\,1.0 & 77  (13) & 0 & 0 & 43.17 & $-$4.47 \\
WDJ171356.07$-$573817.42 & 17230  (330) & 7.88  (0.03) & 0.55  (0.01) & $<$\,1.0 & 97  (13) & 1 & 1 & $-$106.07 & $-$3.67 \\
WD1249+182 & 19960  (480) & 7.87  (0.03) & 0.55  (0.02) & $<$\,1.0 & 48  (9) & 1 & 1 & $-$61.46 & $-$30.67 \\
PG0821+633 & 17060  (360) & 7.88  (0.03) & 0.55  (0.02) & $<$\,1.0 & 101  (16) & 0 & 0 & 33.89 & $-$9.39 \\
WD1325$-$089 & 17180  (370) & 7.88  (0.03) & 0.55  (0.02) & $<$\,1.0 & 98  (15) & 1 & 0 & $-$28.36 & $-$37.04 \\
WD0308+188 & 18180  (370) & 7.87  (0.03) & 0.55  (0.01) & $<$\,1.0 & 75  (12) & 0 & 0 & 63.41 & $-$22.13 \\
WD1535+293 & 23660  (630) & 7.84  (0.04) & 0.55  (0.02) & $<$\,1.0 & 21  (3) & 1 & 1 & $-$17.44 & 27.11 \\
WD1323$-$514 & 18770  (380) & 7.87  (0.03) & 0.55  (0.01) & $<$\,1.0 & 64  (10) & 0 & 0 & $-$154.18 & 9.1 \\
WD0059+257 & 19390  (590) & 7.86  (0.04) & 0.55  (0.02) & $<$\,1.0 & 54  (13) & 1 & 1 & 45.24 & $-$58.74 \\
WD2051+095 & 15080  (340) & 7.88  (0.04) & 0.55  (0.02) & $<$\,1.0 & 161  (23) & 0 & 0 & 10.86 & 26.05 \\
WD1013+256 & 20360  (800) & 7.85  (0.06) & 0.55  (0.03) & $<$\,1.0 & 42  (14) & 1 & 1 & 50.05 & $-$25.53 \\
WD1412$-$109 & 23360  (760) & 7.84  (0.05) & 0.55  (0.03) & $<$\,1.0 & 22  (5) & 1 & 1 & $-$54.17 & $-$25.41 \\
WD1619+123 & 16980  (330) & 7.87  (0.03) & 0.55  (0.01) & $<$\,1.0 & 100  (13) & 0 & 0 & $-$55.66 & $-$9.75 \\
WDJ045514.63$-$544145.41 & 17360  (330) & 7.87  (0.02) & 0.55  (0.01) & $<$\,1.0 & 91  (12) & 0 & 0 & 33.47 & 4.55 \\
WDJ173835.66+415231.19 & 21110  (490) & 7.85  (0.03) & 0.54  (0.02) & $<$\,1.0 & 35  (6) & 1 & 1 & 47.37 & 53.99 \\
WD0114$-$605 & 24440  (550) & 7.83  (0.03) & 0.54  (0.02) & $<$\,1.0 & 18  (2) & 1 & 0 & $-$71.31 & 41.2 \\
PG0816+297 & 16570  (500) & 7.87  (0.04) & 0.54  (0.02) & $<$\,1.0 & 110  (23) & 0 & 0 & 100.55 & $-$6.87 \\
WDJ144107.40$-$560154.83 & 20690  (490) & 7.84  (0.03) & 0.54  (0.02) & $<$\,1.0 & 38  (7) & 1 & 1 & $-$91.56 & 5.59 \\
WD1614$-$128 & 16100  (330) & 7.86  (0.03) & 0.54  (0.02) & $<$\,1.0 & 122  (16) & 0 & 0 & $-$128.85 & $-$20.47 \\
WDJ034835.04+515019.24 & 18570  (410) & 7.84  (0.03) & 0.54  (0.02) & $<$\,1.0 & 64  (11) & 1 & 1 & 60.54 & 6.67 \\
WD0124$-$257 & 22960  (740) & 7.82  (0.06) & 0.53  (0.03) & $<$\,1.0 & 23  (5) & 0 & 0 & 128.46 & 42.92 \\
WD0047$-$524 & 18510  (360) & 7.84  (0.03) & 0.53  (0.01) & $<$\,1.0 & 64  (9) & 0 & 0 & $-$50.36 & 36.19 \\
APASSJ090028.59$-$090923.2 & 20430  (430) & 7.83  (0.03) & 0.53  (0.01) & $<$\,1.0 & 39  (6) & 1 & 0 & $-$8.58 & $-$22.33 \\
PG1128+565 & 26360  (1250) & 7.79  (0.08) & 0.53  (0.04) & $<$\,1.0 & 13  (3) & 0 & 0 & 5.66 & 123.37 \\
HE0452$-$3444 & 20520  (550) & 7.82  (0.04) & 0.53  (0.02) & $<$\,1.0 & 38  (7) & 1 & 0 & 134.53 & 114.58 \\
HS0200+2449 & 18660  (650) & 7.83  (0.05) & 0.53  (0.03) & $<$\,1.0 & 61  (17) & 0 & 0 & 37.23 & $-$9.68 \\
HE0305$-$1145 & 25600  (860) & 7.78  (0.06) & 0.52  (0.03) & $<$\,1.0 & 14  (2) & 0 & 0 & 84.72 & 34.3 \\
WD0839+231 & 25670  (520) & 7.78  (0.03) & 0.52  (0.02) & $<$\,1.0 & 14  (1) & 1 & 1 & 75.23 & $-$16.45 \\
APASSJ151754.65+103043.7 & 18640  (720) & 7.81  (0.06) & 0.52  (0.03) & $<$\,1.0 & 59  (17) & 0 & 0 & $-$91.06 & $-$31.12 \\
PG0846+558 & 26740  (880) & 7.76  (0.06) & 0.52  (0.03) & $<$\,1.0 & 12  (2) & 0 & 0 & 61.56 & 36.69 \\
WDJ180354.33$-$375202.95 & 16480  (610) & 7.81  (0.05) & 0.52  (0.03) & $<$\,1.0 & 102  (25) & 0 & 0 & $-$104.76 & 21.49 \\
WDJ023349.11$-$071534.01 & 21440  (760) & 7.78  (0.05) & 0.52  (0.02) & $<$\,1.0 & 29  (7) & 1 & 1 & 108.72 & $-$17.75 \\
WDJ133752.09+363733.44 & 20410  (440) & 7.78  (0.03) & 0.51  (0.01) & $<$\,1.0 & 37  (5) & 1 & 1 & $-$33.57 & 60.97 \\
WD1932$-$136 & 16650  (440) & 7.8  (0.04) & 0.51  (0.02) & $<$\,1.0 & 95  (18) & 0 & 0 & $-$110.84 & $-$6.33 \\
WDJ161053.38+114352.71 & 19120  (530) & 7.78  (0.04) & 0.51  (0.02) & $<$\,1.0 & 49  (10) & 1 & 0 & $-$45.51 & $-$3.54 \\
WDJ121238.09$-$364240.22 & 19740  (430) & 7.78  (0.03) & 0.51  (0.02) & $<$\,1.0 & 42  (7) & 0 & 0 & $-$58.32 & 6.15 \\
WD0843+516 & 21210  (550) & 7.77  (0.04) & 0.51  (0.02) & $<$\,1.0 & 30  (5) & 1 & 1 & 106.86 & 13.89 \\
PG1620+260 & 27980  (700) & 7.72  (0.04) & 0.5  (0.02) & $<$\,1.0 & 10  (1) & 0 & 0 & $-$124.89 & 43.41 \\
WDJ180240.42$-$243603.86 & 17080  (450) & 7.78  (0.04) & 0.5  (0.02) & $<$\,1.0 & 81  (15) & 1 & 0 & $-$96.14 & $-$10.14 \\
\label{table}
\end{longtable}


\clearpage

\bibliography{bibliography}{}
\bibliographystyle{aasjournal}

\end{CJK*}
\end{document}